\documentclass[preprints,review,accept,moreauthors,pdftex,10pt,a4paper]{mdpi} 

\newcommand\simlt{\lower.5ex\hbox{$\; \buildrel < \over \sim \;$}}
\newcommand\simgt{\lower.5ex\hbox{$\; \buildrel > \over \sim \;$}}

\firstpage{1} 
\pubvolume{xx}
\issuenum{1}
\articlenumber{1}
\pubyear{2019}
\copyrightyear{2019}
\externaleditor{Academic Editor: name}
\history{Received: date; Accepted: date; Published: date}

\pdfoutput=1

\Title{An Introduction to Particle Acceleration in Shearing Flows}

\Author{Frank M. Rieger $^{1,2}$\orcidA{}}

\AuthorNames{Frank M. Rieger}

\address{%
$^{1}$ \quad ZAH, Institut f\"ur Theoretische Astrophysik, Heidelberg University, Philosophenweg 12, 
69120 Heidelberg, Germany; f.rieger@uni-heidelberg.de\\
$^{2}$ \quad Max-Planck-Institut f\"ur Kernphysik, P.O. Box 103980, 69029 Heidelberg, Germany}

\abstract{Shear flows are ubiquitously present in space and astrophysical plasmas. This paper highlights the central
idea of the non-thermal acceleration of charged particles in shearing flows and reviews some of the recent developments. 
Topics include the acceleration of charged particles by microscopic instabilities in collisionless relativistic shear flows, 
Fermi-type particle acceleration in macroscopic, gradual and non-gradual shear flows, as well as shear particle 
acceleration by large-scale velocity turbulence. When put in the context of jetted astrophysical sources such as Active 
Galactic Nuclei, the results illustrate a variety of means beyond conventional diffusive shock acceleration by which 
power-law like particle distributions might be generated. This suggests that relativistic shear flows can account for 
efficient in-situ acceleration of energetic electrons and be of relevance for the production of extreme cosmic rays.}


\begin{document}

\section{Introduction}\label{intro}

Shear flows are naturally expected in a variety of astrophysical environments. Prominent examples include the 
rotating accretion flows around compact objects and the relativistic outflows (jets) in gamma-ray bursts (GRBs) 
or Active Galactic Nuclei (AGN) \citep{Rieger2004}. On conceptual grounds the jets in AGN are expected to 
exhibit some internal velocity stratification from the very beginning, with a black hole ergo-spheric driven, highly 
relativistic (electron-positron) flow surrounded by a slower moving (electron-proton dominated) wind from the inner 
parts of the disk (e.g., see Refs.~\citep{Marti2019,Fendt2019} for recent overviews). In addition, as these jets 
continue to propagate, interactions with the ambient medium are likely to excite instabilities and induce mass 
loading, resulting in a velocity-sheared structure (cf. Ref.~\citep{Perucho2019} for a recent review). 
Radio observations of parcec-scale AGN jets indeed provide evidence for some shear layer morphology, such 
as a boundary layer with parallel magnetic fields or a limb-brightened structure \citep[e.g.,][]{Giroletti2004,
Pushkarev2005,Giroletti2008,Blasi2013,Piner2014,Nagai2014,Gabuzda2014,Boccardi2016}. In connection 
with this, various relativistic hydrodynamic and magneto-hydrodynamic simulations of two-component 
(spine-sheath/layer) and rotating AGN-type jets have been carried out to study their stability properties 
\citep[e.g.,][]{Hardee2003, Meliani2007,Meliani2009,Hardee2007,Mizuno2007,Mizuno2012,Millas2017}, 
indicating for example that the presence of a sheath has a stabilising effect on the jet \citep[e.g.,][]{Hardee2003, Meliani2007,Meliani2009,Hardee2007,Mizuno2007,Mizuno2012,Millas2017}. 
All this suggests that a transversal velocity stratification is a generic feature of AGN-type jets. Given the challenges and
complexity of observed emission properties, this has lead to a renewed interest in multi-zone/spine-shear layer 
acceleration and emission models \citep[e.g.,][]{Aloy2008,Rieger2009,Sahayanathan2009,Tammi2009,Liang2013a,
Laing2014,Tavecchio2014,Tavecchio2015,Rieger2016,Liang2017,Chhotray2017,Liu2017,Kimura2018,Webb2018}.\\ 
One particularly interesting example concerns the emission properties of large-scale AGN jets. Spatially, the fast 
jets of powerful AGN are observed to extend over several hundreds of kilo-parsec (kpc), with bright hot spots being 
formed and significant backflows induced when these jets eventually terminate in the intergalactic medium. Though 
these jets are associated with large fluid Reynolds numbers, they often appear laminar (see e.g. 
Ref.~\citep{Worrall2006} for general orientation). The detection of extended (i.e. kpc-scale), non-thermal X-ray 
emission along several of them indicates that they must contain highly energetic particles \citep{Harris2006,
Rieger2007,Georg2016}. The favoured electron synchrotron explanation in fact implies the presence of 
ultra-relativistic electrons with particle Lorentz factors up to $\gamma~\sim10^8$ \citep[e.g.,][]{Sun2018}. Since
 the typical cooling length of these electrons is very short ($ \ll 1$ kpc), a distributed or continuous (re-)acceleration 
mechanism is required to keep them energized throughout the jet. Stochastic-shear particle acceleration in a
stratified jet has been proposed as possible candidate for this \citep[e.g.,][]{Liu2017}. It seems in principle 
conceivable that particle acceleration in transversal shear flows could also facilitate the acceleration of cosmic rays 
(CRs) to extreme energies \citep[e.g.,][]{Ostrowski1998,Rieger2004,Kimura2018,Webb2018}, supporting the idea 
that large-scale AGN jets are potential ultra-high-energy (UHE)CR acceleration sites \citep[e.g.,][]{Aharonian2002,
Kotera2011}. Since the backflow speeds in AGN can be substantial \citep[e.g.,][]{Perucho2007,Rossi2008,
Matthews2019,Perucho2019b}, this could becomes particularly interesting for cosmic-ray particles that are 
able to sample the velocity contrast between the main jet and its backflow \citep{Webb2018}.\\
The present paper focuses on the potential of fast shearing flows to facilitate efficient particle acceleration. 
As shown below, the transport and acceleration of charged particles in shearing flows can be described and 
explored on different scales, from the plasma skin depth (electron inertial-scale) \citep[e.g.,][]{Alves2012,
Liang2013b} (see Sec.~\ref{skin_depth}) via the relativistic gyro-scale \citep[e.g.,][]{Berezhko1981,Earl1988,
Rieger2007} (see Sec.~\ref{gyro_scale}) to large turbulent length scales \citep{Ohira2013,Lemoine2019} 
(see Sec.~\ref{large_scale}). This paper aims at an accessible introduction to them, reviewing some of the key 
findings along with some recent developments. A particular attention is given to the second one in order to 
recapture early ideas and developments over the years \citep{Berezhko1981,Berezhko1981b,Berezhko1982,
Earl1988,Webb1989,Jokipii1990,Ostrowski1990,Webb1994,Rieger2002,Ostrowski1998,Rieger2004,Rieger2005,
Rieger2006,Rieger2007,Dempsey2009,Rieger2016,Liu2017,Kimura2018,Webb2018}. When viewed in context,
this review shows that a variety of processes beyond conventional diffusive shock acceleration could contribute 
to the efficient energization of particles in jetted astrophysical sources.\\

\section{Supra-thermal Particle Acceleration in Microscopic Shear Flows}\label{skin_depth}
Within recent years, Particle-in-Cell (PIC) simulations have been used to explore the kinetic physics of collisionless, 
un-magnetized, strong (on scales of the electron plasma skin depth) shear flows for different plasma compositions 
(i.e., pure electron-proton, pure electron-positron and some hybrid version) \citep[e.g.,][]{Alves2012,Alves2014,
Alves2015,Grismayer2013,Liang2013a,Liang2013b,Liang2017,Liang2018,Nishikawa2014}. In these simulations a 
variety of microscopic instabilities near the shear surface are encountered that can generate microscopic turbulence 
(i.e., scatterings sites for particle acceleration) as well as lead to the emergence of ordered steady (DC) 
electromagnetic fields and the production of non-thermal particle distributions.

\begin{figure}[htb]
\begin{center}
\includegraphics[width=0.64 \textwidth]{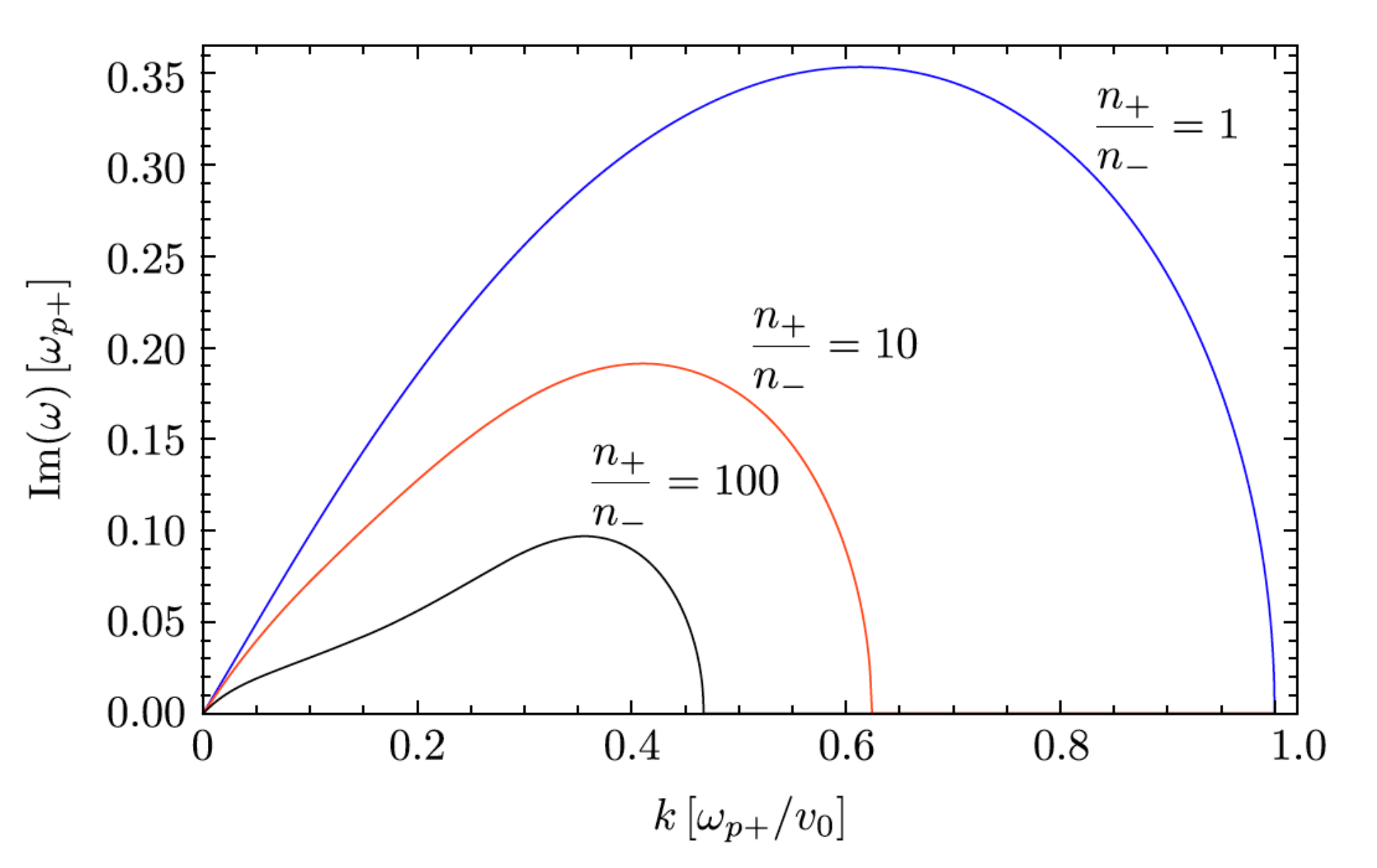}
\caption{Growth rate $Im(\omega)$ of unstable kinetic Kelvin-Helmholtz modes in an electron-proton plasma 
following linear theory for a microscopic tangential shear flow with step-functional velocity jump across the $x
=0$ plane, corresponding to two counter-propagating flows with speed $v_0$ and different densities $n_+$ 
(for the flow in the upper region $x>0$) and $n_-$ (for the flow in the lower region $x<0$). The curves shown 
are for a density contrast $n_+/n_-=1, 10, 100$, respectively. The growth rates display a cut-off at $k\leq 1$ (in 
units of $\omega_{p+}/[v_0 \gamma_0]$,  where $\omega_{p+} = [4\pi n_+ e^2/\gamma_0 m_e]^{1/2}$), and 
take on a maximum value somewhat below. The three curves qualitatively resemble each other, but with a 
growth rate that is lowered for $n_+/n_- >1$. Note, however, that the results of linear fluid theory do not predict
a growth of a DC ($k=0$) mode as found in PIC simulations (see below). From Ref.~\cite{Alves2014}.}
\label{fig1}
\end{center}
\end{figure}

\begin{figure}[htb]
\begin{center}
\includegraphics[width=0.90 \textwidth]{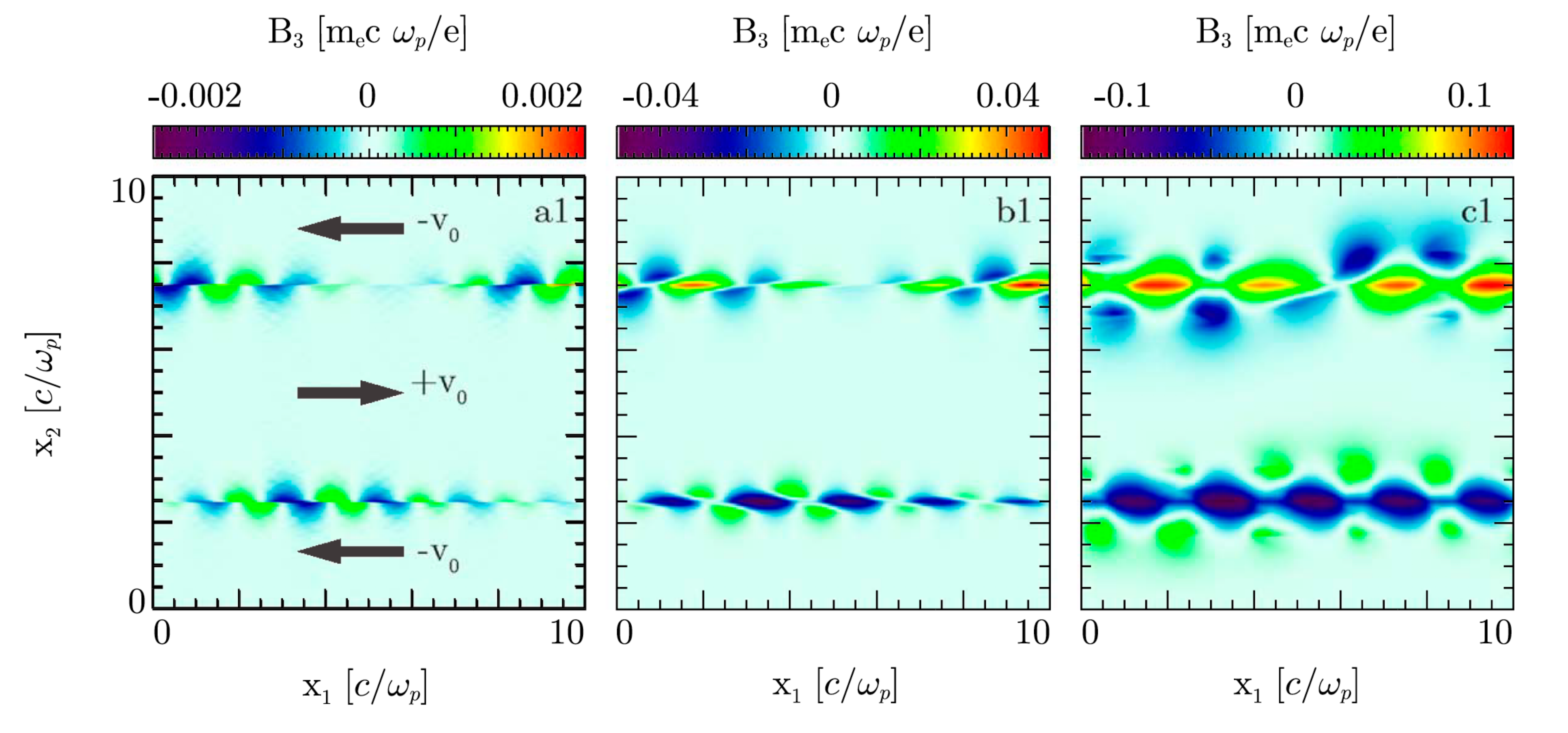}
\caption{Results of two-dimensional PIC simulations of counter-propagating flows of equal density ($n_+=n_-$) 
and with $|v_0|=0.2$ c, showing the emergence of a dominant DC ($k_1=0$ ) magnetic field component along 
the shear on top of the harmonic structure inferred from the linear fluid regime at times $\omega_{pe} t=35$ (a1), 
$45$ (b1) and $55$ (c1). From Ref.~\cite{Alves2014}.}
\label{fig2}
\end{center}
\end{figure}

One well-studied case relates to the excitation of the kinetic (electron-scale) Kelvin-Helmholtz instability (kKHI)
in an un-magnetized electron-proton shear flow which has been explored by means of theory and simulations 
\cite[e.g.,][]{Alves2012,Grismayer2013,Grismayer2013b,Alves2014}. 
The theoretical analysis of the longitudinal kKHI dispersion relation relies on the relativistic fluid formalism of 
un-magnetized plasmas (in the cold plasma limit) coupled with Maxwell's equations, i.e. in cgs units
\begin{eqnarray}
\frac{\partial \rho}{\partial t} + \nabla\cdot \vec{j} = 0 \;,\\
\frac{\partial \vec{p}}{\partial t} + (\vec{v}\cdot\nabla) \vec{p} = 
- e \left (\vec{E} +\frac{\vec{p}}{{\gamma m_e c}}\times \vec{B}\right) \;,\\
\nabla \times \vec{E} = - \frac{1}{c} \frac{\partial \vec{B}}{\partial t}\;,\\
\nabla\times \vec{B} = \frac{4\pi}{c} \vec{j}+\frac{1}{c}\frac{\partial \vec{E}}{\partial t}\;,
\end{eqnarray} where eqs.~(1)-(4) refer to the continuity equation and the conservation of momentum equation, 
respectively, as well as Faraday's and Ampere's law. Here, $\rho:=e\, n$ with $n$ the plasma number density, 
and $\vec{j}, \vec{B}, \vec{E}$ denote the current density, the electric field vector and the magnetic field vector, 
respectively. 
$\vec{p}=\gamma m_e \vec{v}$ and $\vec{v}$ are the linear momentum and velocity of the flow, with $\gamma =
1/(1-v^2/c^2)^{1/2}$ its Lorentz factor. The protons are considered to be free streaming in this approximation.
In order to gain insights into the system one can then specify a simple two-dimensional velocity shear profile,  
\begin{equation}
\vec{v} = v_0(x)\,\vec{e}_y\,,
\end{equation} characterizing a flow that propagates in the $y$-direction with a speed $v_0$ that depends 
on the $x$-coordinate, and prescribe a related (not necessarily constant) density profile $n=n_0(x)$. To study 
the response of the system to perturbations, each of the quantities ($n ,\vec{v}, \vec{E}, \vec{B}, \vec{j}$) are 
further written as $Q(x,y,t)=Q_0(x)+P_1(x,y,t)$ following the usual perturbation analysis (first-order approximation), 
and these expressions then substituted in eqs.~(1)-(4). The resultant equations are then linearized assuming 
that the perturbed quantities scale as 
\begin{equation}\label{pertubation}
P_1(x,y,t) = P_1(x)\, e^{i (k\,y-\omega t)}\,,
\end{equation} with $k\equiv k_y$ parallel to the flow direction. Assuming that external fields are absent ($E_0
=B_0=0$), one can in this way derive a wave equation describing the linear eigenmodes of the system  
(see \citep{Grismayer2013b,Alves2014}), e.g., 
\begin{equation}\label{wave_eq}
\frac{\partial}{\partial x} \left[A \frac{\partial E_{1,y}}{\partial x}\right] + B \frac{\partial E_{1,y}}{\partial x} + C E_{1,y} 
= 0   
\end{equation} for the perturbations $\vec{E}_1$ of the electric field, where $A, B, C$ are functions depending 
on the wave frequency $\omega$, the wave number $k$, the flow velocity $v_0$ and Lorentz factor $\gamma_0
=1/(1-v_0^2)^{1/2}$ as well as on the relativistic plasma frequency $\omega_p(x):= \omega_{pe}(x)/\gamma_0^{1/2}$, 
where $\omega_{pe}$ denotes the usual electron plasma frequency
\begin{equation}
\omega_{pe}(x) = [4\pi n_0(x)\, e^2/m_e]^{1/2} \,.
\end{equation} 
Solutions of eq.~(\ref{wave_eq}) for which $\omega$ becomes imaginary describe unstable modes that grow with
time, cf. eq.~(\ref{pertubation}). Figure~\ref{fig1} provides exemplary solutions for the growth rate of the resultant 
kinetic (electron-scale) Kelvin-Helmholtz instability in the case of a simple tangential shear flow profile 
(i.e., $\vec{v_0}(x)= v_0 \vec{e}_y$ for $x>0$, and $\vec{v_0}(x)=-v_0 \vec{e}_y$ for $x<0$) with different density 
contrast.
The highest growth rate of $\simeq 0.35\, \omega_{p+}/\gamma_0$ is achieved for $n_+/n_- = 1$ at $k_{\rm max}
\simeq 0.61\, \omega_{p+}/(v_0 \gamma_0)$. Note that these modes are obtained in the cold plasma limit, i.e. 
when the effects of thermal motion are negligible compared with $v_0$. Comparing theory and simulation, 
PIC simulations provide general confirmation to this picture, but at late time also reveal the formation of a 
sub-equipartition ($\epsilon_B/\epsilon_p \sim m_e/2 m_p$), large-scale DC ($k=0$) magnetic field component 
along the shear surface with transverse width $\sqrt{\gamma_0} c/\omega_{pe}$ and strength exceeding the 
one at $k_{\rm max}$ by a factor of some few, that is not predicted in linear fluid theory, see Fig.~\ref{fig2}. 
The growth of this DC field cannot be really captured in a fluid description, but seems to be of intrinsically kinetic 
nature. It can be attributed to an effective current density associated with electron transport (mixing) across the 
shear interface. 
As the DC magnetic field grows, the Larmor radius of electrons crossing the shear decreases, until they eventually 
become trapped, imposing an upper limit on the current. In the absence of a large-scale magnetic field the inferred
saturation level of the emergent DC electric and magnetic fields are typically of the order \cite{Grismayer2013}
\begin{equation}
E_{DC}, B_{DC} \sim \sqrt{\gamma_0} m_e c \,\omega_{pe} /e
\end{equation} and seemingly persistent beyond the electron time scale (up to $10^3/\omega_{pe}$). On much 
longer time scales ($> 10^3/\omega_{pe}$), the emergent magnetic fields in two-dimensional PIC simulations 
start to affect the protons more due to their larger gyro-radii, leading to charge separation and the formation of 
a double-layered structure \citep{Liang2013b}.

\begin{figure}[htb]
\begin{center}
\includegraphics[width=0.58 \textwidth]{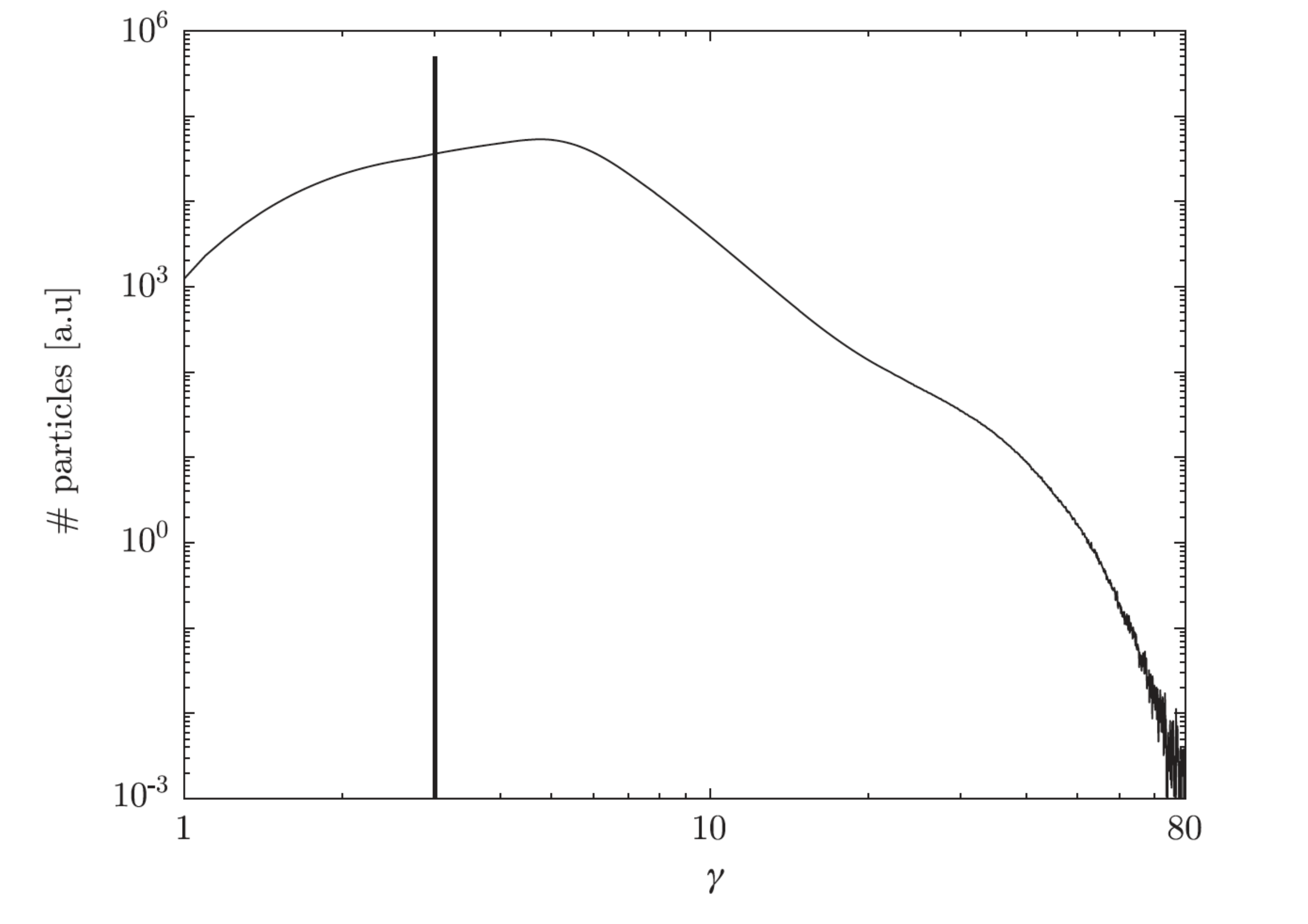}
\includegraphics[width=0.40 \textwidth]{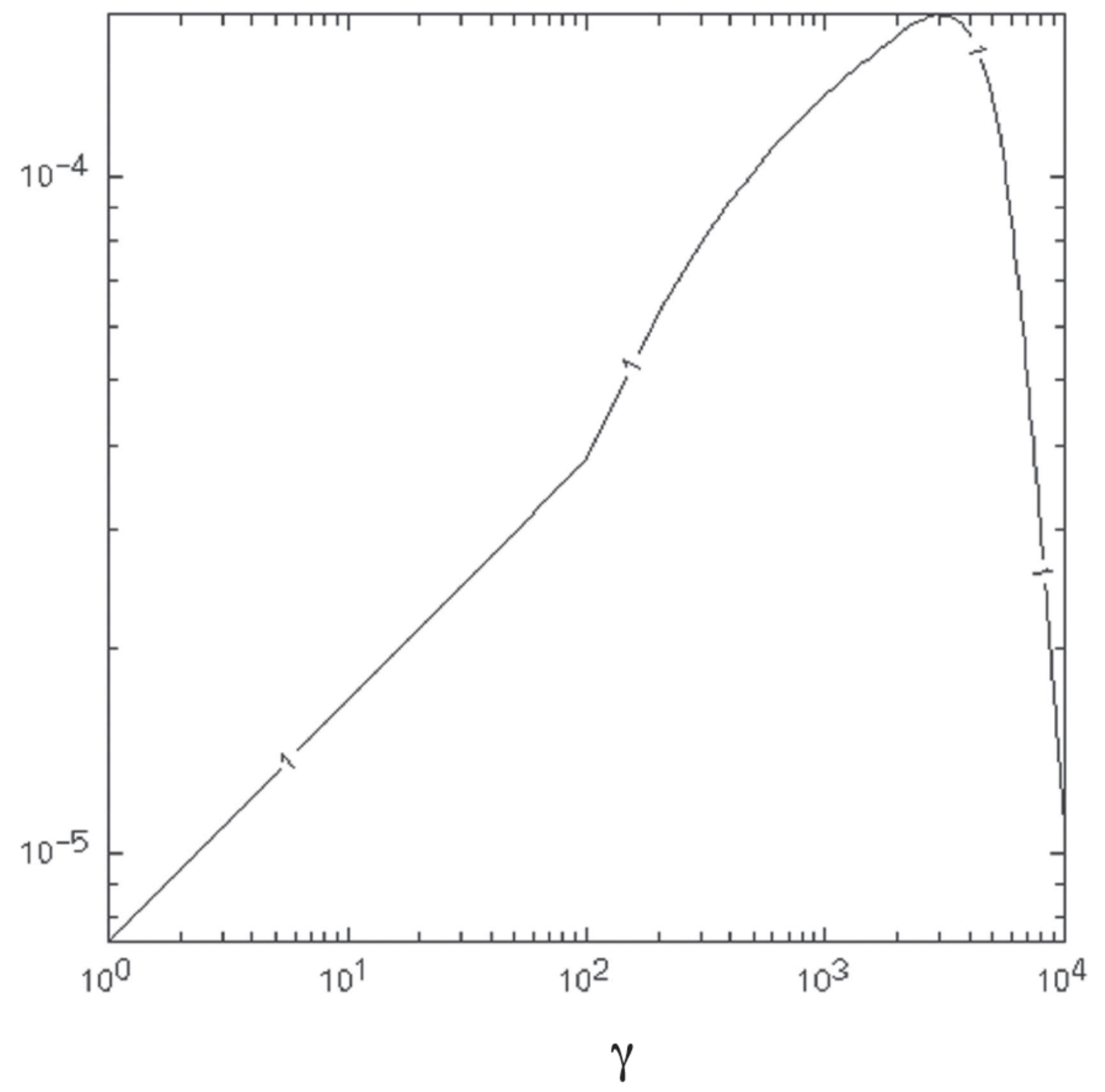}
\caption{{\bf Left:} Resultant electron energy distribution for a relativistic, cold electron-proton shear flow with 
$\gamma_0=3$ at time $t=10^3/\omega_{pe}$. At low energies ($1<\gamma_e<5$) the distributions resembles 
a thermal one, while in the intermediate range $\gamma_e =5-25$ it exhibits some smooth (power-law-type 
$\sim \gamma_e^{-5}$) evolution. After some hardening, a tail is seen extending up to $\gamma_e \sim \gamma_0^4 
\simeq 80$. From Ref.~\citep{Alves2014}.
{\bf Right:} Electron energy distribution in the case of a relativistic shear flow with $\gamma_0=5$ at late times 
$t=10^4/\omega_{pe}$. The distribution peaks around $\gamma_e\sim 3000$, but shows little evidence for a 
conventional power-law behaviour. Both distributions refer to the center-of-momentum frame. From Ref.~\citep{Liang2018}.}
\label{fig3}
\end{center}
\end{figure}

The emergence of organized, self-generated electric and magnetic fields in the shear region could in principle 
lead to efficient particle acceleration. Figure~\ref{fig3} (left) shows an exemplary outcome of a two-dimensional 
PIC simulation of two counter-propagating flows in which electrons experience acceleration in the self-generated 
electric fields along the shear. According to this simulation, electrons might be accelerated up to $\gamma_0^4$ 
times their rest mass energy, possibly with some power-type behaviour in between \citep[][]{Alves2014}.
Other two-dimensional simulations find however, that at late time the electron distribution (at least for higher 
$\gamma_0\geq 5$) reveals a rather narrow peak around the decelerated proton kinetic energy, $\gamma_e 
m_e c^2 \sim (\gamma_0-1) m_p c^2/2$, with a relatively sharp cutoff and little evidence for some power-law 
extension \cite{Liang2018}. 
It has been conjecture that in hybrid positron-electron-ion (instead of a pure electron-proton) plasmas a power-law 
tail could develop due to the presence of non-linear electromagnetic waves facilitating stochastic acceleration of 
electrons \citep{Liang2013b}, though the slopes are expected to be sensitive to the composition and the shear 
flow Lorentz factor $\gamma_0$ \citep{Liang2018}. As things are, the situation appears inconclusive and the 
possible formation of a persistent, long-range power-law electron distribution in microscopic shear flows still 
remains to be demonstrated.

While conceptually highly interesting, the generalisation of these simulation results to realistic astrophysical sources 
is not yet straightforward. The outflows in AGN are likely to be magnetized from the onset, affecting its plasma 
dynamics and instability conditions. It appears conceivable, for example, that magnetic trapping of electrons could 
suppress the growth of a DC field along the shear. While we also know from 3D PIC simulations of non-relativistic,
magnetized plasmas that turbulence can be generated at shear boundary layers \cite[e.g.,][]{Nakamura2013,
Daughton2014,Nakamura2017}, extension to the relativistic regime still remains to be carried out.
In addition, the simulation setups rely on the existence of an idealized (relativistic) velocity jump (shear) on scales 
of the electron skin depth $l_s = c/\omega_{pe}$. For large-scale AGN jets, for example, one typically estimates 
$l_s \simlt 10^{11}$ cm, which would be minute compared to the jet width. Increasing the shear gradient length in 
the noted PIC simulations, however, decreases the growth rate of the kKHI instability \cite[e.g.,][]{Alves2014}, so 
that it appears uncertain to which extent the noted effects should be expected. 
Nevertheless, the simulations mentioned above clearly demonstrate that given suitable conditions microscopic 
instabilities in collisionless relativistic shear flows can efficiently generate electron-scale electromagnetic turbulence, 
allowing for the dissipation of kinetic energy of the flow and the production of supra-thermal particles.

\section{Fermi-type Particle Acceleration in Macroscopic Shear Flows}\label{gyro_scale}
In a seminal paper, E. Fermi \citep{Fermi1949} has proposed a mechanisms for the acceleration of energetic 
charged particles in astrophysical plasma that has become a benchmark for acceleration theory. Fermi-type 
particle acceleration essentially relies on repeated interactions with moving scattering centers,  and typically 
considers this process to be mediated by resonant wave-particle interactions (where the particle mean free 
path $\lambda$ is comparable to the wavelength of the electromagnetic turbulence). In principle, the acceleration 
of energetic particles (of velocity $v\simeq c$) in macroscopic shear flows can be 
understood as such a stochastic process in which particle energization occurs as a results of elastically scattering 
off differently moving (magnetic) inhomogeneities \citep{Rieger2007,Lemoine2019}. In macroscopic shear particle 
acceleration the scattering centers are taken to be frozen into a background flow whose velocity varies with 
transversal coordinate. Their velocities are thus essentially characterized by the general bulk flow profile. 
Depending on the characteristics of the velocity shear, a gradual (continuous) and a non-gradual or discontinuous 
($\lambda >$ characteristic length scale of the velocity shear) case might be distinguished, see Sec.~\ref{gradual}
and \ref{non-gradual} below.\\
The basic concept might be understood with reference to the energy change in an elastic scattering event 
\citep[cf.][]{Rieger2019}. Transforming into the scattering frame and again back to the laboratory frame, a simple 
analysis shows this energy change to be given by 
\begin{equation}\label{energy_change}
\Delta \epsilon = \epsilon_2-\epsilon_1 =  2 
\gamma_u^2 \,(\epsilon_1\, u^2/c^2 - \vec{p}_1\cdot \vec{u})\,, 
\end{equation} where $\vec{u}$ is the characteristic scattering center speed ($u$ its magnitude, $\gamma_u$ the 
corresponding Lorentz factor). $\epsilon_1$ and $\vec{p}_1$ are the initial particle energy and momentum, respectively.  
According to eq.~(\ref{energy_change}) an energetic particle (for which $\epsilon \simeq pc$) will gain energy in 
a head-on (for which $\vec{p}_1\cdot \vec{u}<1$) collision, and lose energy in a following (for which $\vec{p}_1\cdot 
\vec{u}<1$) collision. If one averages over an isotropic particle distribution, however, a net energy gain (i.e. stochastic 
acceleration) is obtained due to the fact that the interaction probability for head-on collisions is higher than the one 
for following collisions. This leads to a second order dependence on $u/c$, such that
\begin{equation}\label{stochastic}
\frac{\left<\Delta \epsilon \right> }{\epsilon}\, \propto \left(\frac{u}{c}\right)^2  > 0\,. 
\end{equation}

\subsection{GRADUAL SHEAR FLOWS}\label{gradual}
\begin{figure}[htb]
\begin{center}
\includegraphics[width=0.50 \textwidth]{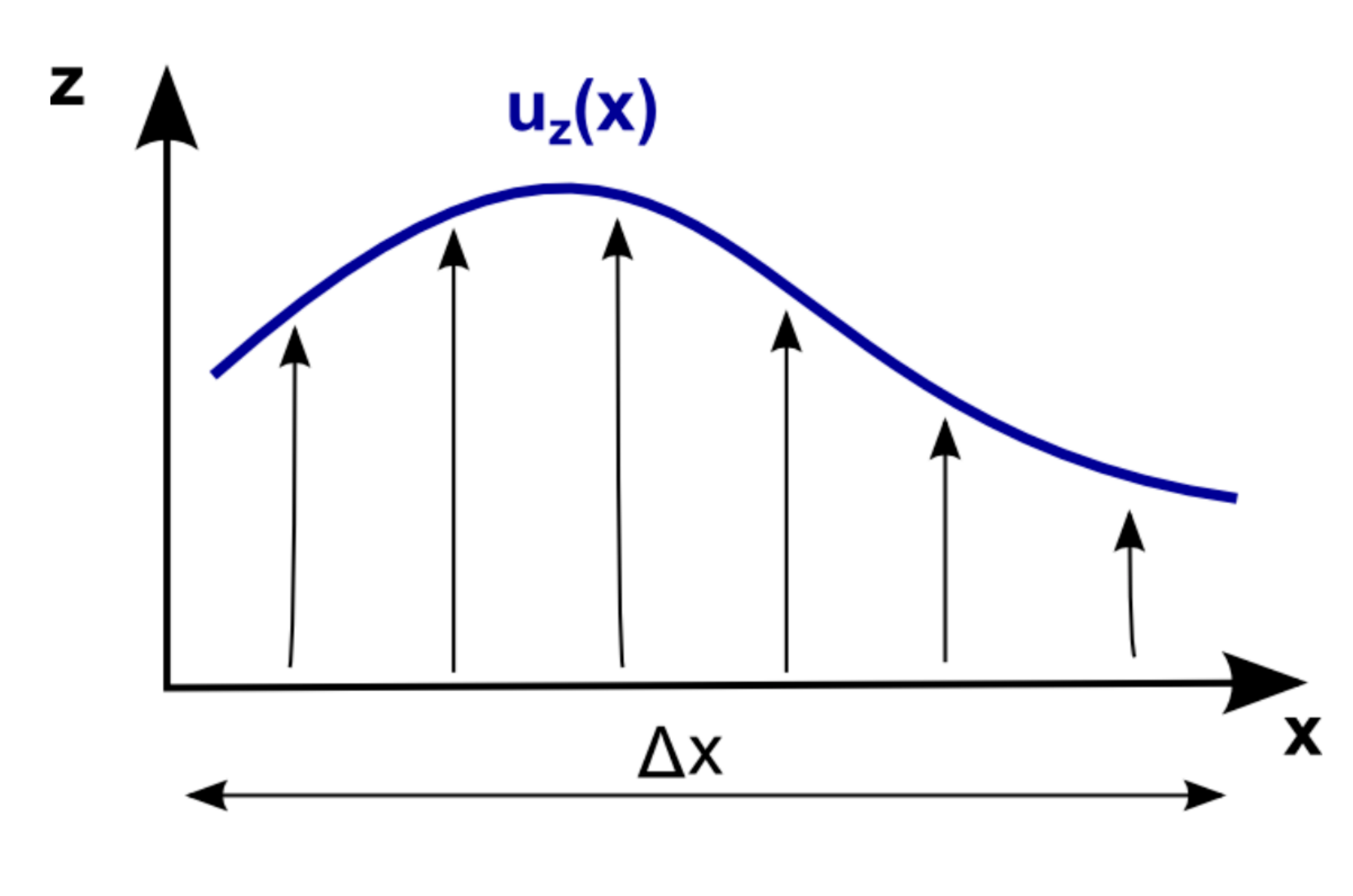}
\caption{Schematic illustration of a simple two-dimensional velocity shear profile, in which a flow, directed 
along the $z$-axis, is characterized by a velocity whose magnitude smoothly varies with the $x$-coordinate. 
$\Delta x$ denotes the shear width.} 
\label{fig4}
\end{center}
\end{figure}

For a particle moving across a continuous non-relativistic shear flow, the flow velocity (and hence the scattering 
center speed) changes by an amount given by the gradient in flow speed, $\nabla u$, multiplied by the particle 
free path, $\lambda=c\,\tau$. Hence, in the context of Fermi-type acceleration, gradual shear particle acceleration 
can be understood as a stochastic acceleration process, in which the conventional scattering center speed is 
replaced by an effective velocity $\bar{u}$ determined by the shear flow profile \citep{Rieger2007} (cf. also 
Ref.~\citep{Rieger2019} for a related review of gradual shear). In the case of a continuous velocity shear $\vec{u} 
= u_z(x) \vec{e}_z$, for example (cf. Fig.~\ref{fig4}), this effective velocity is approximately given by $\bar{u} = 
(\partial u_z/\partial x)\, \lambda$. 
\newline
Accordingly, the fractional energy changes becomes (cf. eq.~[\ref{stochastic}])
\begin{equation}
\frac{\left< \Delta \epsilon \right>}{\epsilon} \propto \left(\frac{\bar{u}}{c}\right)^2 
                                                                    \propto \left( \frac{\partial u_z}{\partial x} \right)^2 
                                                                    \lambda^2.
\end{equation} This suggests a scaling of the characteristic acceleration timescale
\begin{equation}\label{tacc_gradual}
t_{\rm acc} = \frac{\epsilon}{(d\epsilon/dt)} 
                    \sim \frac{\epsilon}{\left<\Delta \epsilon \right>} \cdot \tau 
                    \sim \frac{\epsilon}{\left< \Delta \epsilon \right>} \cdot \frac{\lambda}{c} 
                       \;\propto\; \frac{1}{\lambda}\,,
\end{equation} which, in contrast to classical first and second-order Fermi acceleration, is inversely depending 
on the particle mean free path \cite{Rieger2006}. The principal reason for this seemingly unusual behaviour is 
related to the fact that as a particle increases its energy ($\epsilon \simeq p c $), and thereby its mean free path 
($\lambda(p) \propto p^{\alpha}, \alpha >0$), a higher effective velocity $\bar{u}$ is experienced. 
This scaling has interesting implications, but also implies that for an efficient acceleration of electrons an injection 
of some pre-accelerated seed particles, either from acceleration at shocks or via classical second-order Fermi 
processes is required \cite{Liu2017}. Alternatively, reconnection or plasma turbulence might facilitate electron seed 
injection \cite[e.g.,][]{Sironi2014,Guo2015,Zhdankin2017}. On the other hand, given their larger mean free paths, 
efficient shear acceleration of protons or ions is typically much more easily achieved.

\subsubsection{A Microscopic Approach -- Momentum Space Diffusion}\label{Fokker}
\begin{figure}[htb]
\begin{center}
\includegraphics[width=0.65 \textwidth]{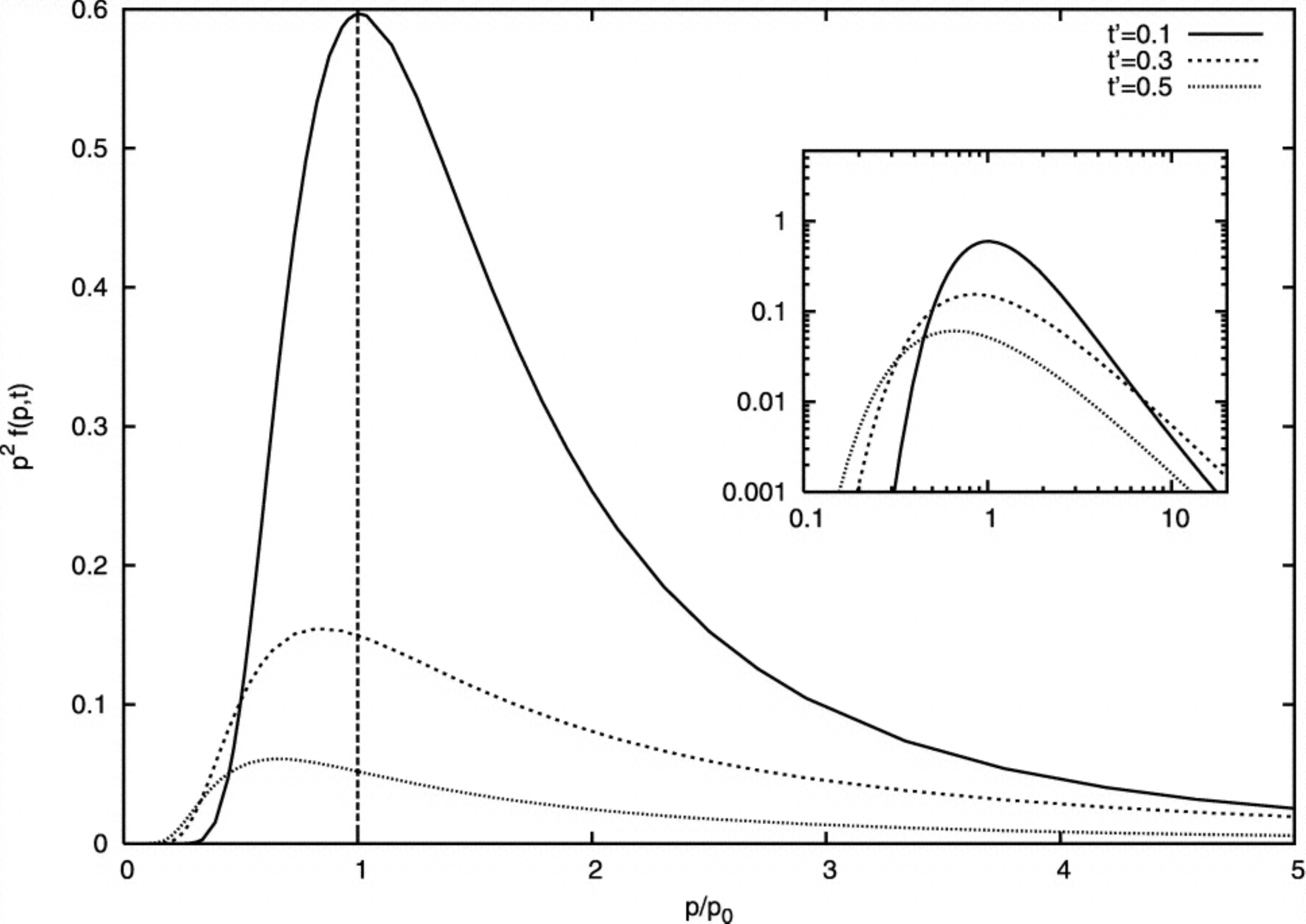}
\caption{Time-dependent solution $f(p,t)$ of the Fokker-Planck diffusion equation for non-relativistic gradual
shear acceleration assuming an impulsive, mono-energetic injection with $p_0$ at $t_0 =0$. 
A linear momentum-dependence $\tau \propto p$ ($\alpha=1$) has been used for the scattering time
The distribution broadens with time due to momentum dispersion. The (double logarithmic) inlet illustrates 
the formation of a power law like distribution $n(p) \propto p^2 f(p) \propto p^{-2}$ above $p_0$ for 
$t'\geq 0.3$. From Ref.~\citep{Rieger2006}.}
\label{fig5}
\end{center}
\end{figure}

As a stochastic process, gradual shear particle acceleration is accompanied not only by a momentum change 
(energy gain), but also by momentum dispersion (broadening) term. This can be shown in a microscopic picture 
by evaluating the average rate of momentum change and dispersion, i.e., by calculating the respective 
Fokker-Planck coefficients. For simplicity, consider again a non-relativistic shear flow with $\vec{u} = u_z(x) \vec{e}_z$.
While travelling for one scattering time $\tau$ across such a flow, the momentum of a particle relative to it changes 
by $\vec{p}_2 = \vec{p}_1+ m\,\delta\vec{u}$, where $\delta\vec{u} = (\partial u_z/\partial x)\, \delta x\, \vec{e}_z$ 
and $\delta x = v_x \,\tau$, with $v_x$ the $x$-component of the particle velocity, and $m$ the relativistic particle
mass. In general, the timescale for collisions (mean scattering time) is expected to be an increasing function of 
momentum, i.e., $\tau \equiv \tau(p) = \tau_0\, p^{\alpha}$. By expanding $\Delta p \equiv (p_2-p_1)$ to second 
order in $\delta u/c$ and averaging over an isotropic particle distribution, the Fokker-Planck coefficients become
\cite{Jokipii1990,Rieger2006}
\begin{eqnarray}
\left<\frac{\Delta p}{\Delta t}\right>  \propto  \frac{<\Delta p>}{\tau}
                                                         \; \propto \;  p \left(\frac{\partial  u_z}{\partial x}\right)^2 \tau\,,
                                                      \nonumber \\
\left<\frac{(\Delta p)^2}{\Delta t}\right>  \propto  \frac{<(\Delta p)^2>}{\tau}
                                                       \; \propto  \; p^2 \left(\frac{\partial  u_z}{\partial x}\right)^2 \tau\,.
\end{eqnarray} One can show that these coefficients are related by the equation
\begin{equation}\label{balance}
\left<\frac{\Delta p}{\Delta t}\right> = \frac{1}{2p^2} \frac{\partial}{\partial p} 
                                                          \left[p^2 \left<\frac{(\Delta p)^2}{\Delta t}\right>\right]
                                                          =\frac{\Gamma}{p^2}\frac{\partial}{\partial p}\left(p^4 \tau\right)\,,
\end{equation} i.e., that they satisfy the principle of detailed balance (scattering being reversible). Here, the 
symbol $\Gamma$ on the right hand side of eq.~(\ref{balance}) denotes the shear flow coefficient, which for 
the employed flow profile is simply given by $\Gamma = (1/15) (\partial u_z/\partial x)^2$.\\
Under the condition of detailed balance, the associated Fokker-Planck equation is known to reduce to a diffusion 
equation in momentum space. Hence in the absence of radiative losses and escape, the momentum-space 
particle distribution $f(p,t)$ experiencing gradual shear acceleration obeys a simple Fokker-Planck 
type diffusion equation \citep{Jokipii1990,Rieger2006}
\begin{equation}\label{FP}
\frac{\partial f(p,t)}{\partial t} = \frac{1}{p^2} \frac{\partial}{\partial p}
                                                \left(p^2 D_{sh}\,\frac{\partial f}{\partial p}\right)\,,
\end{equation} where $D_{sh}:=\Gamma p^2 \tau$ denotes the momentum space (shear) diffusion coefficient.
For the considered application
\begin{equation}\label{Dsh}
D_{sh} = \Gamma p^2 \tau = \frac{1}{15} \left(\frac{\partial u_z}{\partial x}\right)^2 p^2 \tau\,,
\end{equation} with $\tau \equiv \tau(p) =  \tau_0 \, p^{\alpha}$. 
Figure~\ref{fig5} shows an exemplary solution of eq.~(\ref{FP}) for the case $\tau \propto p$ \cite{Rieger2006}.
At sufficiently large times, the particle distribution above injection approaches a power law shape 
\begin{equation}\label{powerlaw}
n(p) \propto p^2 f(p) \propto p^{-(1+\alpha)}\,
\end{equation} for $\alpha >0$, with power law index depending on the momentum scaling of the particle mean 
free path ($\lambda \simeq c \tau \propto p^{\alpha}$) \citep{Berezhko1981,Berezhko1982,Rieger2006}. 
In particular, for a gyro-dependence, $\alpha =1$, one has $n(p) \propto p^{-2}$, which is comparable to the
index for first-order Fermi acceleration at non-relativistic (high Mach number) shocks. The quasi steady-state 
(time-integrated) particle distribution $f(p)$ for continuous injection then becomes constant below $p_0$, and 
takes on the noted power-law shape $f(p) \propto p^{-(3+\alpha)}$ above it.  

\subsubsection{Propagation and Acceleration in Non-relativistic Shear Flows}
In order to take spatial transport into account, a suitable particle transport equation has to be derived. For 
general shear flows this requires an extension of the original Parker transport equation \citep{Parker1965}, 
describing the propagation and acceleration of energetic particles scattered by turbulent inhomogeneities 
embedded in a background plasma. In the limit of non-relativistic flow speeds this has been done by Earl, 
Jokipii \& Morfill~\citep{Earl1988}, starting from the non-relativistic Boltzmann equation for the phase-space 
distribution $f(\vec{x},\vec{p},t)$ and assuming a simple BKG-type collision term, ($\partial f/\partial t)_s = 
(f - \langle f \rangle)/\tau$. The approach utilises a mixed system of phase-space coordinates such that 
quantities which are operated upon the scattering operator, i.e. the momentum, are evaluated in the 
co-moving flow frame (allowing for a convenient treatment of the scattering physics), while time and space 
coordinates are still measured in the laboratory frame. Given non-relativistic flow speeds, the particle 
momentum components are then related by a Galilean transformation $p_i = p_i'+ m\,u_i$, where $u_i$ 
denotes the background flow speed and $p_i'$ the co-moving particle momentum. The analysis 
subsequently only requires that scattering is strong enough to guarantee the diffusion approximation, i.e., 
to ensure that departures of the particle distribution from isotropy are small, hence that the particle 
distribution function is well approximated by $f = f_0 + f_1$, with $<f_1> = 0$ and $f_1\ll f_0$. The full 
non-relativistic particle transport equation for $f_0(\vec{x}, p',t)$ eventually takes the form \citep{Earl1988}
\begin{eqnarray}\label{transport}
   \!\!\!\frac{\partial f_0}{\partial t} 
   \!\!&+\!\!&  u_i\,\frac{\partial f_0}{\partial x_i}
   -\frac{p'}{3}\,\frac{\partial u_i}{\partial x_i}\,\frac{\partial f_0}{\partial p'}
   -\frac{\partial}{\partial x_i}\left(\kappa\,\frac{\partial f_0}{\partial x_i}\right)
   + \frac{2\,\tau\,p'}{3}\,A_i\,\frac{\partial^2 f_0}{\partial x_i \partial p'} \nonumber \\
   \!\!&+&\!\! \frac{1}{3\,p'^2}\,\frac{\partial\,(\tau\,p'^3)}{\partial p'}\,A_i\, \frac{\partial f_0}{\partial x_i}
   - \frac{\Gamma}{p'^2}\,\frac{\partial}{\partial p'}\left(\tau\,p'^4\,
   \frac{\partial f_0}{\partial p'}\right) + \frac{p'}{3}\,\frac{\partial (\tau A_i)}{\partial x_i}\,
   \frac{\partial f_0}{\partial p'} = 0\,,
  \end{eqnarray}
  ($i=1,2,3$). Here, $\kappa$ denotes the isotropic spatial diffusion coefficient, $\kappa(p) = \tau(p) v^2/3
  \simeq \tau(p) c^2/3$, $A_i$ is the total acceleration vector
  \begin{equation}\label{accel}
  A_i:=\frac{D u_i}{Dt} = \frac{\partial u_i}{\partial t} + u_l\,\frac{\partial u_i}{\partial x_l}\,,
  \end{equation} and $\Gamma$ is the viscous shear flow coefficient given by
  \begin{equation}\label{nr_shear_coefficient}
  \Gamma=\frac{1}{30}\left(\frac{\partial u_i}{\partial x_k}
  + \frac{\partial u_k}{\partial x_i}\right)^2 
  -\,\,\frac{2}{45}\,\frac{\partial u_i}{\partial x_i}\,
  \frac{\partial u_k}{\partial x_k}\,.
\end{equation}  
The second, third and fourth term in eq.~(\ref{transport}) describe the well-known effects of convection, adiabatic 
energy change and spatial diffusion \citep{Parker1965}, while the terms involving $A_i$ describe the effects of 
inertial drifts (cf. also \citep{Williams1991,Williams1993} for incorporation of a mean magnetic field). The additional 
term involving $\Gamma$ characterizes energy changes due to flow shear and divergence.\\ 
When a steady (non-relativistic) shear flow of the form $\vec{u} = u_z(x) \vec{e}_z$ is considered, the adiabatic 
and inertial terms vanish ($\partial u_i/\partial x_i=0$ and $A_i =0$), while $\Gamma$ becomes $\Gamma = (1/15) 
(\partial u_z/\partial x)^2$. The space-independent part of eq.~(\ref{transport}) then reduces to 
\begin{equation}\label{non_rel_transport}
\frac{\partial f_0}{\partial t}= \frac{\Gamma}{p'^2}\,\frac{\partial}{\partial p'}\left(\tau\,p'^4\,
   \frac{\partial f_0}{\partial p'}\right)\,,
\end{equation} which coincides with eq.~(\ref{FP}). For completeness, let us mention, that a significant velocity shear 
could also occur during magnetic reconnection, suggesting that shear acceleration in outflowing regions can contribute 
to particle energization if the particle distribution is sufficiently anisotropic \cite[e.g.,][]{leRoux2015,leRoux2018,Li2018}.

\subsubsection{Generalization of the Particle Transport to Relativistic Shear Flows}
In order to compete with the diffusive escape of particles, efficient shear particle acceleration generally requires relativistic 
flow speeds (see also below). This then demands a suitable extension of the particle transport equation to the relativistic 
regime as has been obtained by Webb et al. \citep{Webb1989,Webb1994}, utilising as before a mixed-frame approach
(with the momentum being evaluated in the co-moving flow frame). Assuming isotropic diffusion with $\kappa$ and denoting 
the (covariant) metric tensor by $g_{\alpha \beta}$, the zero component of the (comoving) particle momentum four vector by 
$p'^0=E'/c$, the fluid four velocity by $u_{\alpha}$ and the fluid four acceleration by $\dot{u}_{\alpha}:= u^{\beta} 
\nabla_{\beta} u_{\alpha}$, where $\nabla_{\beta} u_{\alpha}$ denotes covariant derivation, the full particle transport 
equation for the isotopic distribution function $f_0(x^{\alpha},p')$ with $x^{\alpha} = (ct,x,y,z,)$ takes the form 
\citep{Webb1989}
\begin{eqnarray}\label{Webb_eq}
&&\nabla_{\alpha} \left[c u^{\alpha} f_0 -  \kappa \left(g^{\alpha \beta}+u^{\alpha}u^{\beta}\right)
                                \left(\frac{\partial f_0}{\partial x^{\beta}} - \dot{u}_{\beta} \frac{(p'^0)^2}{p'}
                                \frac{\partial f_0}{\partial p'}\right)\right]\nonumber \\
             && + \frac{1}{p'^2} \frac{\partial}{\partial p'} \left[-\frac{p'^3}{3} c\,\nabla_{\beta} u^{\beta} f_0
                            +p'^3 \left(\frac{p'^0}{p'}\right)^2 \kappa \,\dot{u}^{\beta}
                             \left(\frac{\partial f_0}{\partial x^{\beta}} - \dot{u}_{\beta} \frac{(p'^0)^2}{p'}
                             \frac{\partial f_0}{\partial p'}\right) - \Gamma \tau p'^4 \frac{\partial f_0}{\partial p'}\right] = Q\,,
                             \nonumber \\    
                             \label{GRPTE}
\end{eqnarray} where $Q$ denotes the source term and Greek indices ($\alpha, \beta$) run from 0 to 3. 
$\Gamma$ denotes the (generalized) relativistic shear coefficient, which in the strong scattering 
limit is given by
\begin{equation}
\Gamma = \frac{c^2}{30} \sigma_{\alpha\beta}\sigma^{\alpha\beta}\,,
\end{equation} where $\sigma_{\alpha\beta}$ is the (covariant) fluid shear tensor given by
\begin{equation}
\sigma_{\,\alpha\,\beta}=\nabla_{\alpha} u_{\beta}+\nabla_{\beta} u_{\alpha}
                 +\dot{u}_{\alpha} u_{\beta}+\dot{u}_{\beta} u_{\alpha}
                 -\frac{2}{3}\left(g_{\,\alpha\,\beta}+u_{\alpha}\,u_{\beta}
                  \right)\,\nabla_{\delta} u^{\delta}\,.
\end{equation}
\newline 
In the case of a cylindrical jet with a steady (relativistic) shear flow profile $\vec{u} = u(r) \vec{e}_z$, the fluid four 
acceleration ($\dot{u}_{\alpha}=0$) and divergence ($\nabla_{\beta} u^{\beta}=0$) vanish, retaining only the shear
term with 
\begin{equation}
\Gamma=(\gamma_u^4/15)\,(du/dr)^2\,
\end{equation} with $\gamma_u(r)=1/(1-[u(r)/c]^2)^{1/2}$, in the second line of eq.~(\ref{GRPTE}). The characteristic 
co-moving acceleration timescale then becomes  \citep[][]{Rieger2004,Webb2018}
\begin{equation}
t_{\rm acc}' = \frac{15}{(4+\alpha)\, \gamma_u^4 \,(du/dr)^2\, \tau'} \,,
\end{equation} where $\tau'\propto p'^{\alpha}$ is the mean scattering time.\\
Full z-independent, steady-state solutions of eq.~(\ref{GRPTE}) for such a flow profile and specific forms of the radial 
dependence of $\kappa(r,p)$ have been recently presented by Webb et al. \cite{Webb2018,Webb2019} and discussed
in connection with extragalactic radio jets. In principle, eq.~(\ref{GRPTE}) also allows to treat particle acceleration in 
relativistic outflows where intrinsic jet rotation introduce a velocity shear. In such a case a complex interplay between 
shear and centrifugal effects can occur \citep{Rieger2002,Dempsey2009}. This could be of particular relevance in the 
context of AGN-type jets where some internal jet rotation is expected \citep[e.g.,][]{Meliani2007,Meliani2009,Millas2017}.\\
Note that when the generalized relativistic transport equation~(\ref{Webb_eq}) is reduced to its non-relativistic limit, 
also an additional term quadratic in the acceleration vector $\dot{u}^i \dot{u}_i \propto (A_i)^2$ is recovered, which as 
such does not appear in the previous version of the non-relativistic transport equation~(\ref{transport}) \cite{Earl1988,
Webb1989,Williams1993}. This is due to the fact that in the derivation of eq.~(\ref{transport}), focusing on cosmic-ray 
transport, the relevant terms have been neglected as they are typically of order $(u/c)^2$ smaller than the viscous term.

\subsubsection{Recent Applications of Gradual Shear Acceleration}
A variety of topics and applications have been discussed in the context of gradual shear acceleration 
\citep[e.g.,][]{Rieger2007,Rieger2009,Sahayanathan2009,Liu2017,Webb2018}. In the following three recent 
results are briefly mentioned:
\begin{itemize}
\item {\it (i) Shear Particle Acceleration in Expanding Relativistic Outflows:}\\
The jetted outflows from AGN and GRBs can exhibit highly relativistic speeds, regions of (quasi-conical) expansion
and flow Lorentz factors varying with polar angle \citep[e.g.,][]{Pushkarev2017,Kovalev2019,Salafia2015,Zhang2018}. 
This makes them possible sites where gradual shear particle acceleration could occur \citep{Rieger2005,Rieger2016}. 
An application to AGN-type outflows has been presented recently, considering the case of a radial velocity shear 
profile $u^{\alpha} = \gamma_b\left(\theta)(1, v_r(\theta)/c,0,0\right)$, where $\theta$ denotes the polar angle, $r$ 
the radial coordinate, and $\gamma_b(\theta)$ the bulk flow Lorentz factor \citep{Rieger2016}. When the impact
of different functional dependencies for $\gamma_v(\theta)$ such as a power-law-, Gaussian- or Fermi-Dirac-type 
profile is explored (see Fig.~\ref{fig6}), the characteristic (co-moving) acceleration timescale is found to be a strong 
function of $\theta$. This could facilitate the generation of some prominent, non-axis (e.g., 'ridge line') emission 
features in AGN jets \citep{Rieger2016}.
\begin{figure}[hbt]
\begin{center}
\hspace*{0.5cm}
\includegraphics[width=0.38 \textwidth]{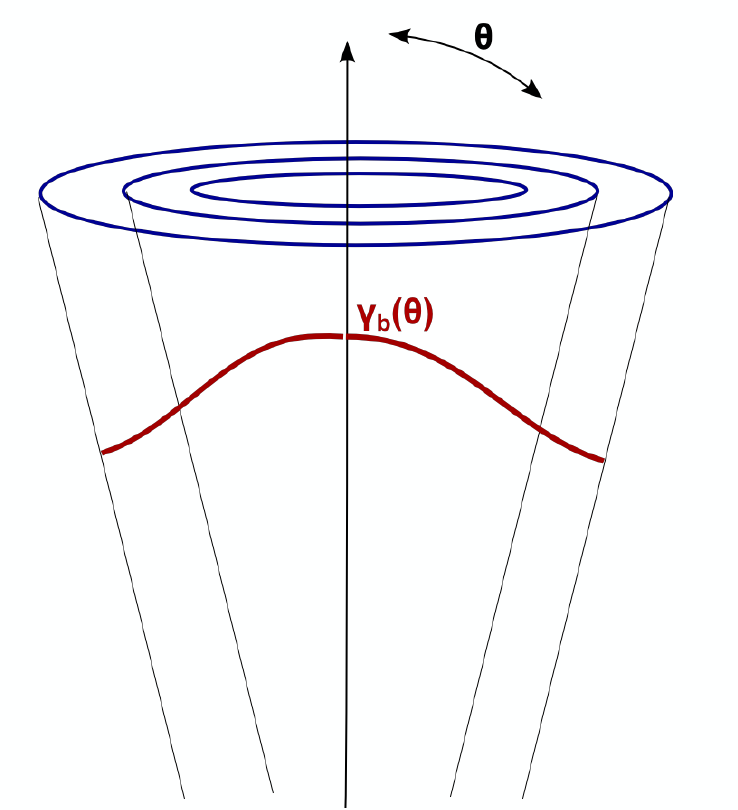}
\includegraphics[width=0.56 \textwidth]{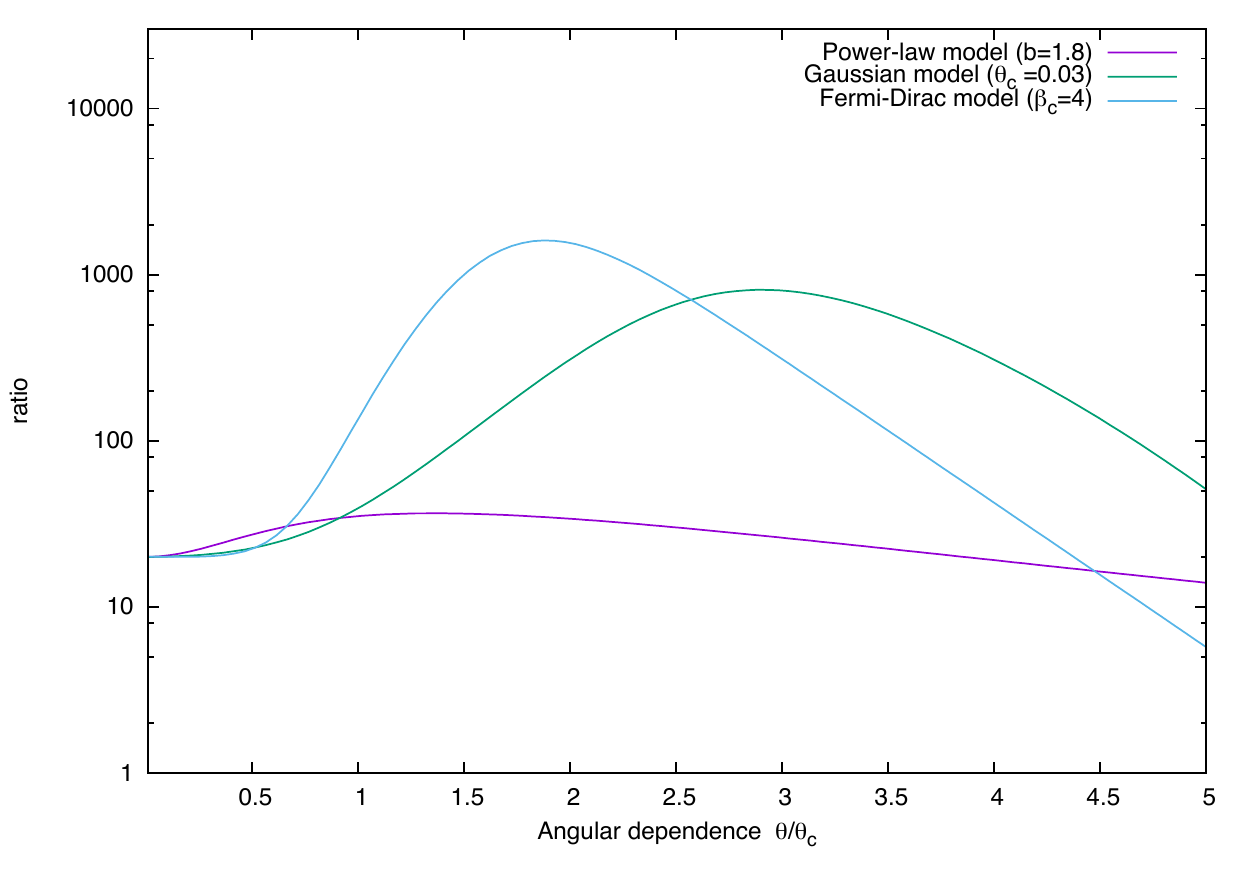}
\caption{{\bf Left:} Illustration of a simple conical flow whose radial outflow speeds varies with polar angle $\theta$.
{\bf Right:} Ratio of viscous shear gain versus adiabatic losses multiplied by ($r/\lambda'$), illustrated assuming a 
core angle $\theta_c =0.03$ rad and an on-axis flow Lorentz factor $\gamma_b=30$. A non-axis preference 
becomes particularly evident for a Gaussian or Fermi-Dirac shaped flow profile. From Ref.~\citep{Rieger2016}.}
\label{fig6}
\end{center}
\end{figure}
In order to overcome adiabatic losses ($\propto \gamma_b v_r/r$) and allow for efficient acceleration, relativistic 
outflow speeds and sufficient energetic seed particles ($\lambda'/r>10^{-3}$ for the example shown in Fig.~\ref{fig6}) 
would be needed. When put in GRB context, particle acceleration in expanding shear flows might result in a weak 
and long-duration leptonic emission component in GRBs, as well as be conducive to UHE cosmic-ray production 
\cite{Rieger2005}.\\
\item {\it (ii) Multi-Component Particle Distributions and Extended Emission:}\\
Since $t_{\rm acc} \propto 1/\lambda$ (eq.~[\ref{tacc_gradual}]), gradual shear particle acceleration will begin to
dominate over conventional first- and second-order Fermi acceleration ($t_{\rm acc} \propto \lambda$) above 
a certain energy threshold. This could naturally result in the formation of multi-component particle distributions. 
A basic example assuming radiative-loss-limited acceleration in a cylindrical, mildly relativistic shearing flow is 
shown in Fig.~\ref{fig7} \cite{Liu2017}. The figure is based on a time-dependent solution of the Fokker-Planck 
equation for $f(p,t)$, or equivalently $f(\gamma,t)$, including the effects of classical second-order Fermi and 
gradual shear particle acceleration as well as synchrotron losses. Employing a Kolmogorov-type ($q=5/3$) scaling
 for the particle mean free path, $\lambda(p) \propto p^{2-q}$, and using parameters applicable to mildly relativistic 
large-scale jets in AGN, electron acceleration up to Lorentz factors of $\gamma \sim10^9$ seems feasible 
(cf. Fig.~\ref{fig7} [left]). In the example given, stochastic second-order Fermi acceleration dominates particle 
energization up to $\gamma \sim 10^4$, while above this threshold shear acceleration becomes operative leading 
to a somewhat flatter spectral slope (with a change by $2/3$ in the example shown). Synchrotron radiation eventually 
introduces a spectral cut-off at high energies.
\begin{figure}[hbt]
\begin{center}
\includegraphics[width=0.53 \textwidth]{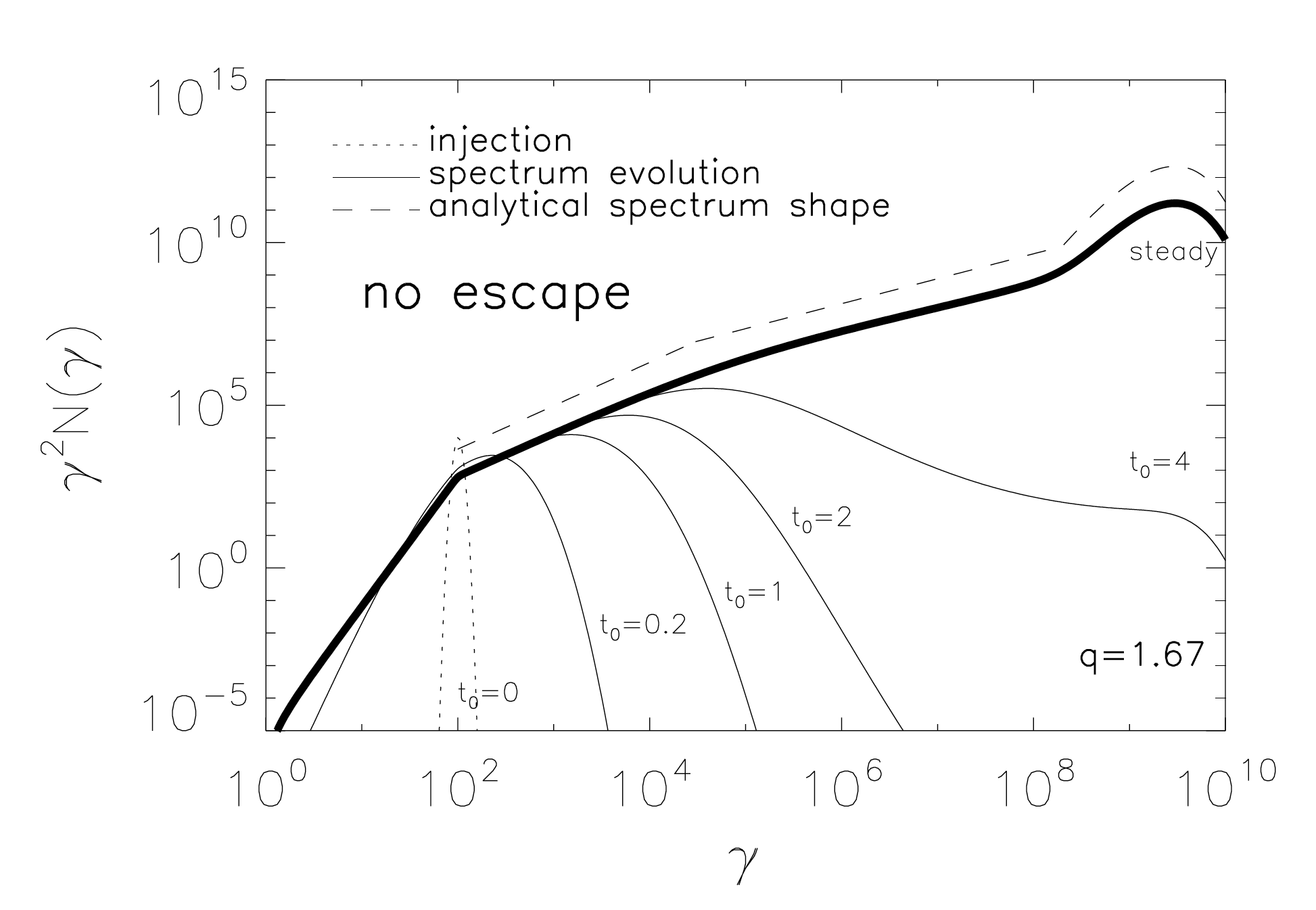}
\includegraphics[width=0.42 \textwidth]{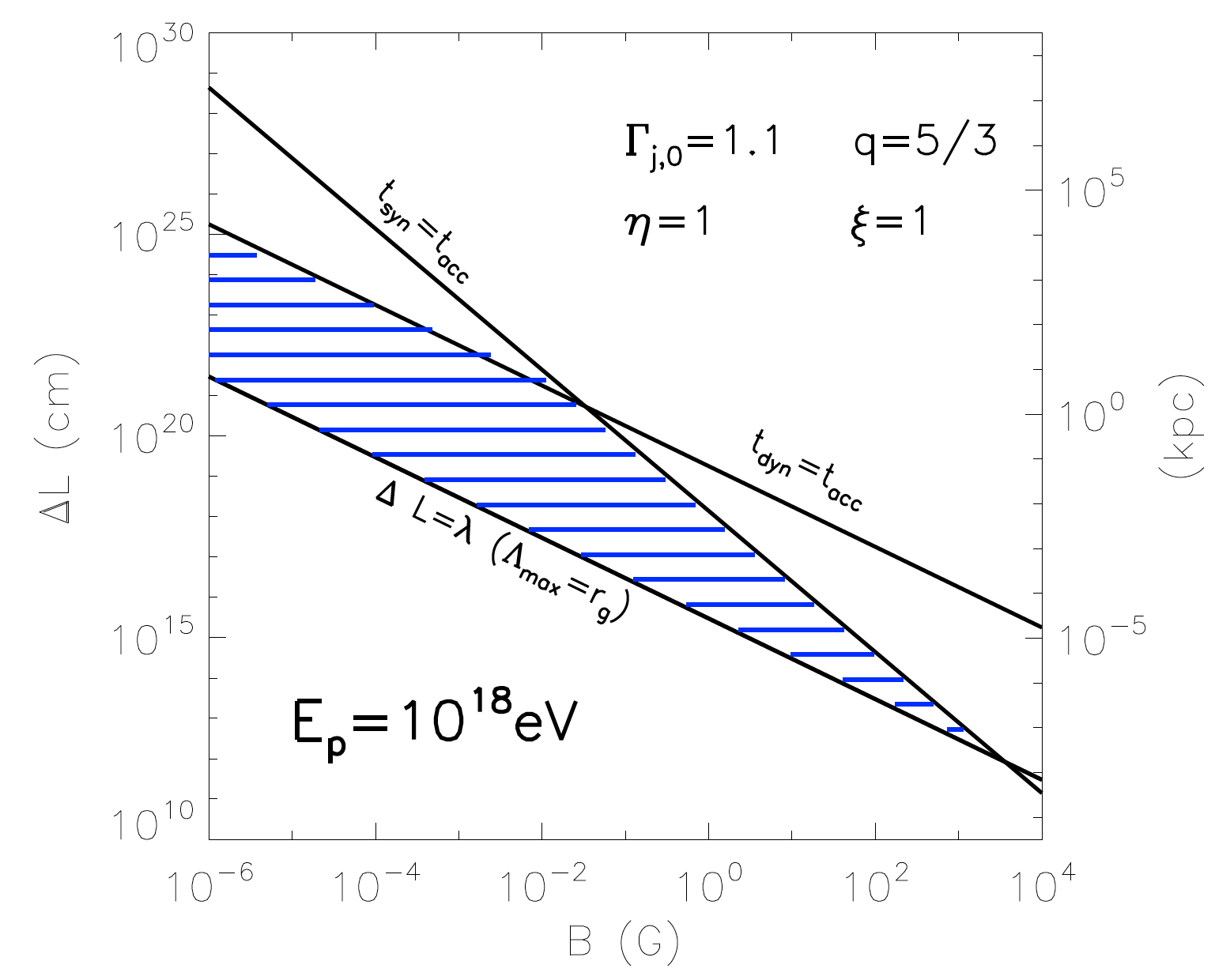}
\caption{{\bf Left:} Time-evolution of the electron spectrum, $\gamma^2\, n(\gamma)$, in the presence of 
stochastic-shear particle acceleration, where $n(\gamma) \propto \gamma^2 f(\gamma)$ represents a solution 
of the corresponding Fokker-Planck equation for a linearly decreasing (trans-relativistic) velocity shear of width 
$\Delta l \sim  r_j/10$, and an Alfven speed $\beta_A  \sim 0.007$. Above particle Lorentz factors of a few times 
$10^4$ the spectrum is shaped by shear acceleration, with a high-energy spectral cut-off around $\gamma\sim
10^9$ being introduced by synchrotron losses. The successive operation of different acceleration processes here 
naturally results in a broken-power law distribution. {\bf Right:} Required (blue-hatched) range of parameters 
(magnetic field strength $B$, shear layer width $\Delta l$) to allow shear acceleration of protons to $\sim10^{18}$ 
eV given confinement and loss constraints for the noted conditions. The required conditions might be met in 
large-scale AGN jets. From Ref.~\citep{Liu2017}.}
\label{fig7}
\end{center}
\end{figure}
As shearing conditions are likely to prevail along astrophysical jets, stochastic-shear particle acceleration is
expected to be of relevance for understanding the extended X-ray emission in the large-scale jets of AGN 
(cf. Sec.~\ref{intro}) \citep{Liu2017}. In reality, the anticipated change in spectral slope will also depend on 
the spatial transport and escape properties (see below). As a consequence, higher speeds would be needed 
to achieve comparable, moderate breaks.
When put in UHE cosmic-ray context, gradual shear acceleration of protons up to $\sim10^{19}$ eV seems 
feasible in the large-scale jets of AGN \citep{Liu2017,Webb2018,Webb2019}, cf. also Fig.~\ref{fig7} (right). 
Higher energies might be achieved for faster flows and for heavier particles.\\
\item {\it (iii) Incorporating Spatial Transport and Diffusive Escape:}\\
In the previous Fokker-Planck approach details of the spatial transport, and possible modifications introduced by 
the diffusive escape of particles from the system, have not been incorporated. Implications of the spatial transport 
could in principle be studied by using the full relativistic particle transport equation~(\ref{GRPTE}). Analytical 
examples in this regard have been recently presented by Webb et al.~\citep{Webb2018,Webb2019}. Focusing on 
steady-state solutions $f_0(r, p')$ for a cylindrical jet with longitudinal shear $u_z(r)$ and allowing for a specific 
radial dependence $g(r)$ of the scattering time, $\tau(r, p) = \tau_0\,g(r)\,(p/p_0)^{\alpha}$, they showed that 
diffusive escape can counter-act efficient acceleration. In particular, while the local particle distribution still follows 
a power law $f(p') \propto p'^{-\mu}$, its momentum index $\mu$ becomes dependent on the maximum flow speed 
$\beta_0$ on the jet axis, and significantly steepens with decreasing $\beta_0$ (approaching $\mu \rightarrow 
\infty$ for $\beta_0\rightarrow 0$) \citep{Webb2018,Webb2019}. Though possible limitations due to the chosen 
$\tau$-dependence may deserve some further studies, these results imply that efficient gradual shear particle 
acceleration requires relativistic flow speeds. The analytical solutions \cite{Webb2018} can be used to explore 
the full radial evolution of the particle transport. Figure~\ref{fig8} represents an example for a hyperbolic, relativistic 
shear flow profile $\beta_z(r)=\beta_0 [1-\tanh(r)^2]$ with a maximum Lorentz factor $\gamma_b=20$ on the jet 
axis \citep{Rieger2019}.
\begin{figure}[h]
\begin{center}
\includegraphics[width=0.55 \textwidth]{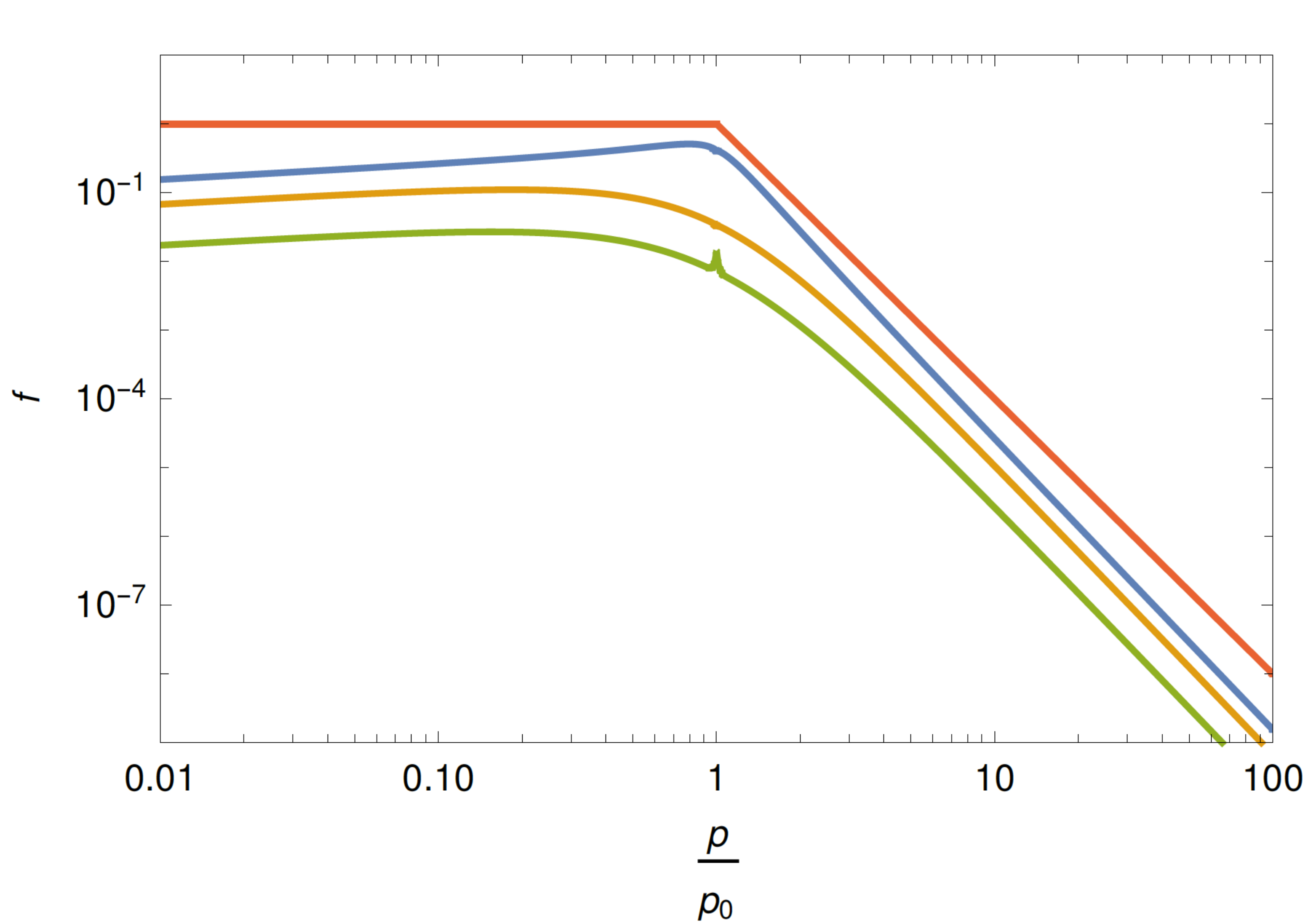}
\caption{Normalized steady-state particle distribution function $f(r,p')$ in the presence of a gradual, hyperbolic 
relativistic shear flow as a function of momentum $(p'/p_0')$, shown for three different spatial locations, $r=0.06$ 
(blue), $r=0.50$ (orange), $r=1.20$ (green). Mono-energetic particle injection with $p_0'$ at $r_1=0.02$ and an 
outer (escape) boundary at $r_2=2$ have been assumed. A linear momentum-dependence $\alpha =1$ has been 
used for the scattering time. The red curve (top) shows the expected power-law dependence $f(p') \propto p'^{-4}$ 
above $p_0'$, as inferred from the Fokker-Planck approach in eq.~(\ref{powerlaw}).}
\label{fig8}
\end{center}
\end{figure}
As can be seen, away from injection at $r_1$ the known power-law momentum dependence, eq.~(\ref{powerlaw}), 
is approximately recovered at high flow speeds ($\beta_0 \rightarrow 1$). Clearly, advancing our understanding 
of the (radial) diffusion properties in astrophysical jets will be important to further improve our understanding of
the particle acceleration in gradual shear flows.
\end{itemize}

\subsection{NON-GRADUAL SHEAR FLOWS}\label{non-gradual}
\begin{figure}[htb]
\begin{center}
\includegraphics[width=0.48 \textwidth]{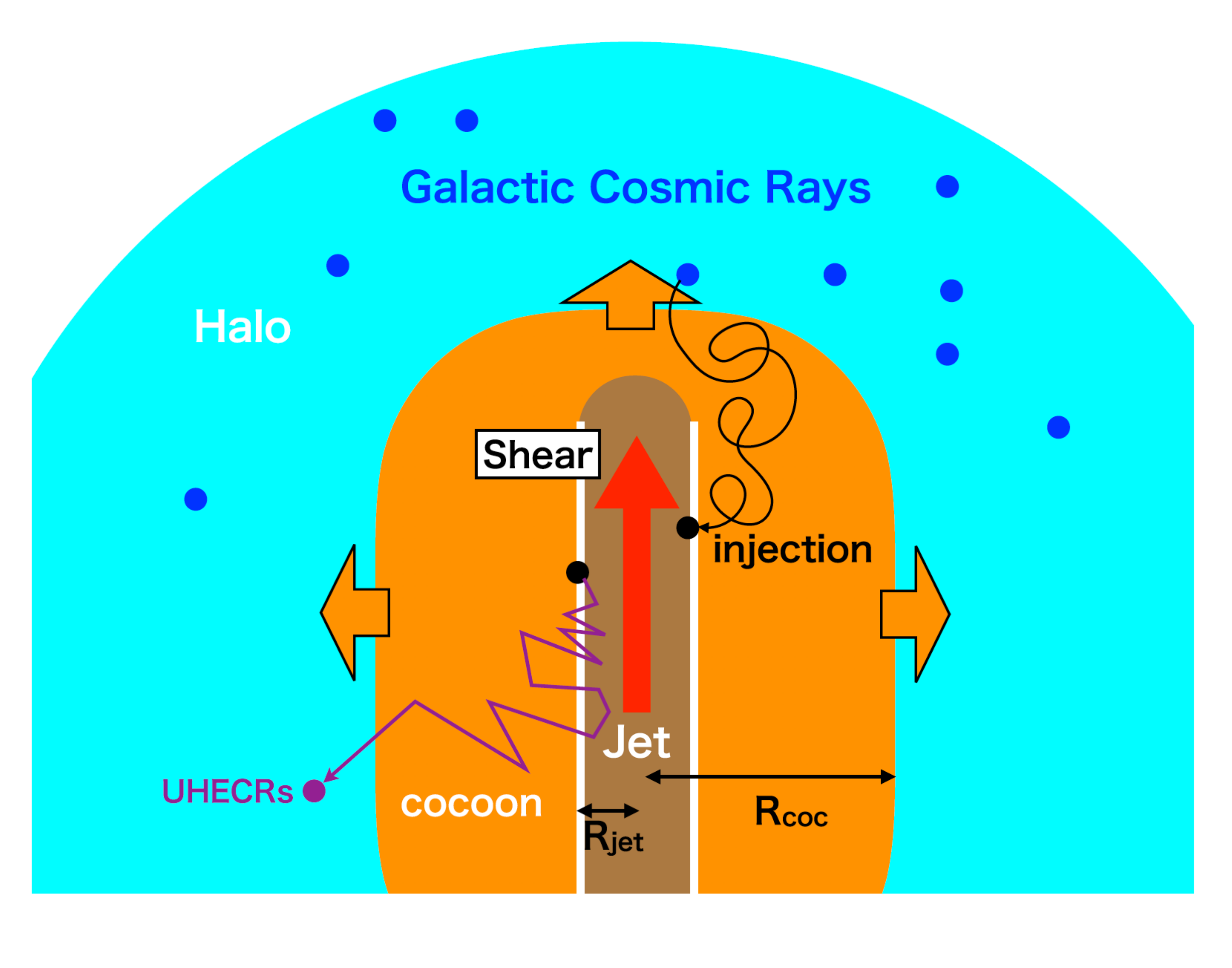}
\includegraphics[width=0.47 \textwidth]{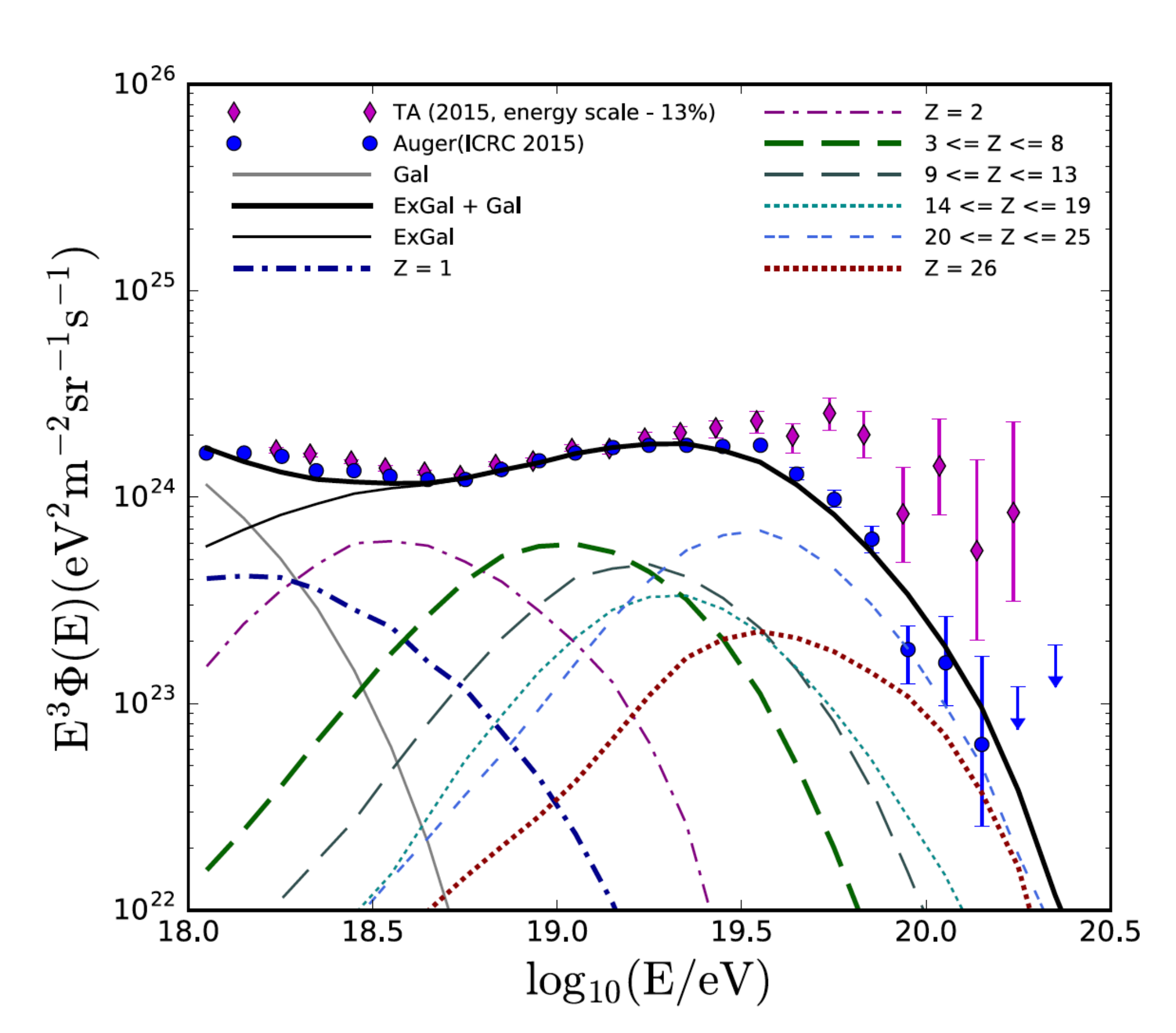}
\caption{{\bf Left:} Cartoon of the considered scenario assuming a recycling of galactic cosmic rays by 
non-gradual shear acceleration in a jet - (turbulent) cocoon system. Some fraction of galactic cosmic 
rays are considered to be swept up by the kiloparsec-scale jet and reaccelerated to high energies. 
The return probability of a particle here is dominated by the scattering (turbulence) properties (i.e.,
particle mean free path) in the cocoon and not in the jet. {\bf Right:} Reconstruction of the observed 
UHECR spectrum assuming mildly relativistic ($\Gamma_j\simeq 1.4$), non-gradual shear 
acceleration in an extragalactic jet-cocoon system with a thin transition layer $\Delta r=5$ pc 
($B_j=0.3$mG, $r_j=0.5$ kpc). The composition at the highest end is dominated by intermediate 
and heavy nuclei. From Ref.~\citep{Kimura2018}.}
\label{fig9}
\end{center}
\end{figure}

Once the particle mean free path becomes larger than the characteristic scale ($\Delta r$) of the shear 
transition layer, acceleration essentially becomes non-gradual. A particle may then be viewed as passing 
almost unaffected through the layer and experiencing a strong, quasi-discontinuous jump in velocity.
Such a situation could arise, for example, at the interface between the ambient medium and the interior 
of powerful (FR II-type) large-scale AGN jets \citep{Ostrowski1990, Ostrowski1998,Ostrowski2000,
Rieger2004}. If a particle is able to repeatedly cross the transition layer, efficient particle acceleration to 
high energy may occur. This may allow to boost pre-accelerated cosmic rays further to ultra-high energies 
(UHECR) \citep[e.g.,][]{Ostrowski1998}, or to enhance the electron synchrotron emission of large-scale 
AGN jets \citep[e.g.,][]{Stawarz2002}. In case of ultra-relativistic jet flow velocities ($\Gamma_j \gg 1$) 
a "one-shot boost" by $\sim \Gamma_j^2 \gg 1$ might occur (crossing and re-crossing the layer), cf. 
eq.~(\ref{energy_change}). It has been argued that this could be sufficient to boost seed galactic cosmic 
rays of energy $\simlt 10^{17}$ eV in blazar-type AGN to ultra-high energies $> 10^{18}$ eV
\citep{Caprioli2015,Caprioli2019}.\\
If the particle distribution would remain nearly isotropic near the shear discontinuity, the mean energy for 
a single crossing is approximately given by $\left< \Delta E/E \right> \simeq (\Gamma_j - 1)$ \citep[e.g.,][]{Rieger2004}.
This would suggest that the increase in particle energy could be substantial provided the velocity shear is 
sufficiently relativistic. For non-relativistic velocities ($\Gamma_j\sim 1$), on the other hand, only the usual 
energy gain of second order in $\Delta u$ is obtained. To properly treat relativistic flow speeds ($\Delta u/c
\simeq 1$) the non-negligible anisotropy of the particle distribution needs to be taken into account. The 
principal effects of this is a reduction in efficiency. Accordingly, the mean energy gain may be expressed
as
\begin{equation}
        \left< \frac{\Delta E}{E}\right> =\eta_e\, (\Gamma_j - 1)\,,
\end{equation} where $\eta_e < 1$. Monte Carlo particle simulations within the strong scattering limit (i.e., 
assuming $\Delta B/B \sim 1$) suggest that $\eta_e$ may still be a substantial fraction of unity
\citep{Ostrowski1990}. One can then define an acceleration timescale $t_{\rm acc} = \tau / \left<\Delta E/E \right>$, 
where $\tau$ is the mean time for boundary crossing and $\tau = \lambda / c$, with $\lambda\sim r_g$ 
the particle mean free path ($r_g$ the gyro-radius) \citep[cf.][]{Rieger2004}. In the laboratory (ambient 
medium) rest frame one thus obtains \citep{Ostrowski1998,Ostrowski2000}
\begin{equation}
   t_{\rm acc} = \alpha \;\, \frac{\lambda}{c}\;\; \simgt \;(1-10) \; \frac{r_g}{c} \quad
                        {\rm provided} \;\; r_g >  \Delta r \,\,.
\end{equation} Simulations suggest that for suitable choices (high $\Delta u$, small $r_{\rm max}$) $\alpha$ 
might be as small as $\sim (1-10)$ assuming that particles are allowed to escape once they have crossed 
a boundary at some lateral distance $r_{\rm max}$. In general, however, $\alpha$ is a sensitive function 
of $r_{\rm max}$, increasing quasi-linearly with increasing $r_{\rm max}$ \citep{Ostrowski1990}. 
Nevertheless, provided particles with $\lambda \sim r_g > \Delta r$ and $r_g < r_j$ ($r_j$ denoting the jet 
radius) are present, acceleration may proceed fairly quickly.\\
Observations of pc-scale AGN jets with evidence for a shear layer morphology (e.g., a boundary layer with 
parallel magnetic fields or limb-brightened structure \citep[e.g.,][]{Giroletti2004,Pushkarev2005,Giroletti2008,
Blasi2013,Piner2014}) suggest that $\Delta r < 0.5\,r_j$. Taking $\Delta r \sim 0.1\,r_j \sim 0.1$ pc and $B 
\sim 0.01$ G for a semi-quantitative estimate, the condition $r_g>\Delta r$ would require very energetic
seed electrons of Lorentz factor $\gamma_e \sim 10^{12} (\Delta r/0.1~\mathrm{pc})$ and protons of Lorentz 
factors $\gamma_p \sim 5 \cdot 10^8  (\Delta r/0.1~\mathrm{pc})$, respectively. The associated acceleration 
timescale would be of the order of $t_{\rm acc} \simgt \Delta r/c \sim 0.3 \, (\Delta r/0.1~\mathrm{pc})$ yrs. 
Hence, unless the transition layer would be much narrower, the considered mechanism may not work 
efficiently for electrons given their rapid synchrotron losses. The mechanism is, however, much more 
favourable for protons (or cosmic rays). Given suitable seed injection (e.g., by gradual shear) cosmic rays 
may be further accelerated until their gyro-radius $r_g$ becomes larger than the width of the jet $r_j$. 
Note that for non-gradual shear one has $t_{\rm acc}\propto \lambda$, while for gradual shear $t_{\rm acc} 
\propto 1/\lambda$ (eq.~[\ref{tacc_gradual}]).\\
An application of mildly relativistic ($\Gamma_j\simeq 1.4$), non-gradual shear acceleration to the possible 
energization of UHE cosmic rays in the kiloparsec-scale jets of FR~I type objects has been recently presented 
by Kimura et al. \citep{Kimura2018}. In the considered setup (Fig.~\ref{fig9}) galactic cosmic rays are swept up 
by the jet and reaccelerated to UHE energies. Monte Carlo simulations suggests that cosmic rays escaping 
through the cocoon exhibit a very hard, power-law like spectrum $dN/dE \propto E^{-1} - E^0$ and a cut-off 
around the maximum energy that is much slower than exponential \citep{Kimura2018}. The results are 
somewhat sensitive to the turbulence description in the cocoon (e.g., coherence scale, cocoon size) and the 
assumed thickness $\Delta r$ of the transition layer (defining the injection energy threshold of galactic cosmic 
rays).
Enlarging $\Delta r$ and considering a strongly turbulent layer (see above), for example, is likely to affect
the results. Nevertheless, these simulations show that non-gradual shear acceleration in large-scale AGN 
jets (not necessarily of the FR~II type) could in principle play an important role in the (re)acceleration of 
UHE cosmic rays.\\

\section{Particle Acceleration by Large-Scale Velocity Turbulence}\label{large_scale}
\begin{figure}[htb]
\begin{center}
\includegraphics[width=0.70 \textwidth]{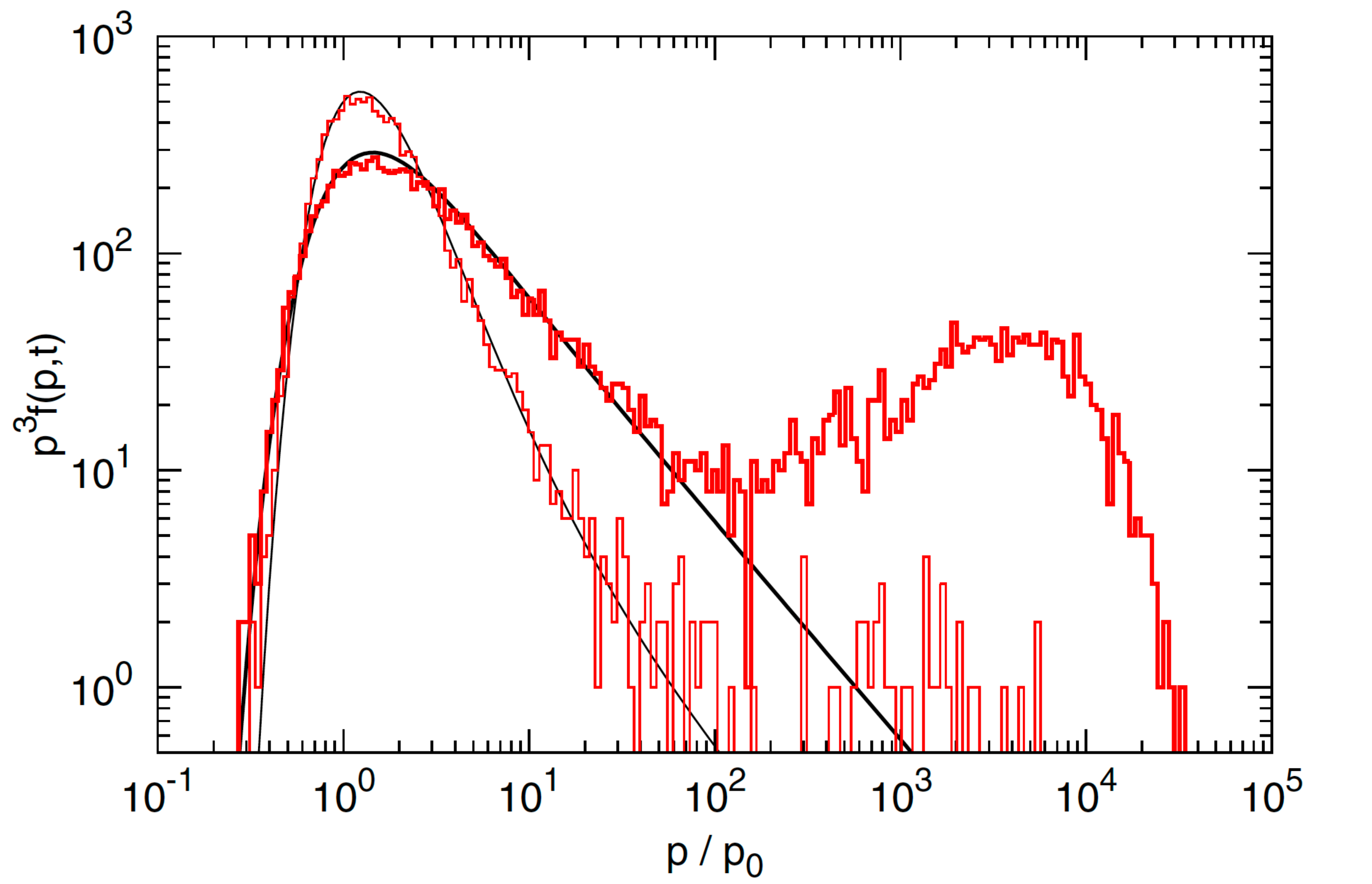}
\caption{Particle distribution function for turbulent shear acceleration at two different times assuming static
incompressible velocity turbulence with mean amplitude $\langle \delta u^2 \rangle =(0.05 c)^2$. Red histograms 
show results of Monte Carlo simulations for a mono-chromatic wave spectrum $S(k)\propto \delta(k-k_0)$ with 
$\tau_0 c k_0=0.01$, where $\tau(p) =\tau_0\,(p/p_0)$ (i.e., $\alpha=1$) has been employed. Thin and thick 
lines show analytical solutions at times $t/\tau_0 = 5\times 10^6$ and $10^7$, respectively. The formation 
of a power-law tail $f(p)\propto p^{-4}$ above injection $p_0$ and below $p/p_0\sim 10^2$ becomes apparent 
with time. Deviations with regard to the analytical solutions are seen towards higher momenta ($p/p_0>10^2$) 
where the particle mean free path $\lambda_{res}$ starts exceeding the turbulence scale $L_0=2\pi/k_0$. 
From Ref.~\citep{Ohira2013}.}
\label{fig10}
\end{center}
\end{figure}

If the turbulence scale of the flow is much larger than the particle mean free path, i.e. $\lambda_{turb} = 2\pi/
k  \gg r_{g}$, stochastic (non-resonant) particle acceleration could in principle occur due to random compression 
and rarefaction/decompression of the medium, or due to incompressible large-scale motions. Particle acceleration 
by large-scale (long-wavelength) {\it compressible} velocity turbulence has been studied some time ago
\cite[e.g.,][]{Bykov1982,Bykov1983,Ptuskin1988}, and more recently discussed with respect to the production 
of supra-thermal ions in the solar wind \citep{Fisk2008,Jokipii2010}. Interestingly however, particle acceleration 
may also occur in {\it incompressible} (divergence-free, $\nabla \cdot \delta\vec{u}=0$) velocity turbulence, i.e. 
by scattering centers (small-scale inhomogeneities) carried by a plasma flow with large-scale velocity fluctuations 
\citep{Bykov1983}, resulting in what has been referred to as "turbulent shear acceleration" (TSA) \citep{Ohira2013}. 
The total energy change for an ensemble of particles will obviously be sensitive to the presumed turbulent velocity 
field, with different descriptions yielding different efficiencies.\\ 
For non-relativistic turbulence (cf. also Ref.~\cite{Lemoine2019} for a discussion of relativistic turbulence), one 
can draw on eq.~(\ref{transport}) to describe the ensemble-averaged particle transport. Its space-independent 
part, or respectively the equation for the spatially and ensemble averaged distribution function $f$, then reduces 
to a diffusion equation in momentum space (cf. also eq.~[\ref{FP}])
\begin{equation}\label{TSA_FP}
\frac{\partial f(p,t)}{\partial t} = \frac{1}{p^2} \frac{\partial}{\partial p}
                                                \left(p^2 D_{TSA}\,\frac{\partial f}{\partial p}\right)\,,
\end{equation} where $D_{TSA}(p)$ is the momentum space (TSA) diffusion coefficient. 
Using eq.~(\ref{nr_shear_coefficient}) and assuming a pure static, homogeneous and isotropic, incompressible 
velocity turbulence, $u_i(t,\vec{x}) = \delta u_i(\vec{x})$, $\left<\delta u_i\right> =0$, this coefficient can be written as 
\citep{Ohira2013}
\begin{equation}\label{DTSA}
D_{TSA}(p) = \frac{2}{15} p^2 \tau \int \frac{d^3k}{(2\pi)^3}\,S(k)\,k^2\,.
\end{equation} Here, $\tau \equiv \tau(p) = \tau_0 \, p^{\alpha}$ again denotes the mean scattering time, and $k$ 
and $S(k)$ are the wavenumber and spectrum of the incompressible velocity turbulence. Equation~(\ref{DTSA})
implies that $D_{TSA}$ is dominated by small-scale (large $k$) turbulence whenever $k^5 S(k)$ is an increasing 
function of $k$. Hence for a (3D) Kolmogorov-type spectrum \citep{Ohira2013}, $S(k) k^2 \propto k^{-5/3}$, i.e.
\begin{equation}
S(k) \propto k^{-11/3}\
\end{equation} in the range $k_0 \leq k \leq k_{\rm max}$, turbulent shear acceleration would become relevant
towards the smaller scales. The upper limit of the $k$-integration in eq.~(\ref{DTSA}) should not exceed $k_{\rm res}$ 
and hence approximately be given by $min\{k_{\rm max}, k_{\rm res}\}$, where $k_{\rm max}$ is the maximum 
wave number of the turbulence, $k_{\rm res} \sim 1/(\tau\,v) \propto  p^{-\alpha}$ and $k_0=2\pi/L_0$ ($L_0$ being 
the turbulence injection scale). For $k_{\rm res}< k_{\rm max}$ the diffusion coefficient for the noted spectrum 
thus scales as 
\begin{equation}\label{DTSA2}
D_{TSA}(p) \propto p^{2+\alpha}\, (k_{\rm res})^{4/3} \propto  p^{2-\alpha/3}\,,
\end{equation} while for $k_{\rm res} > k_{\rm max}$, or for a mono-chromatic spectrum $S(k) \propto \delta(k-k_0)$, 
one instead obtains $D_{TSA}(p) \propto p^{2+\alpha}$. Analytical solutions of eq.~(\ref{TSA_FP}) for general 
momentum indices of $D_{TSA}$ can be found in, e.g. Ref.~\citep{Rieger2006}. 
Monte Carlo simulations of the acceleration of particles in static, homogeneous and isotropic incompressible 
turbulence have been presented by Ohira~\citep{Ohira2013}, confirming the general picture. Figure~\ref{fig10}
provides an illustration for turbulent shear particle acceleration in mono-chromatic turbulence. 
The simulation results are in good agreement with analytical expectations up to $p/p_0>10^2$, where in the 
chosen setup the particle mean free path starts to exceed the turbulence scale ($\lambda_{\rm res}(p) > 
L_0$) and approximations for the analytical treatment no longer apply. \\ 
The characteristic acceleration timescale for non-relativistic turbulent shear for the case of $k_{\rm res}<
k_{\rm max}$ (eq.~[\ref{DTSA2}]) is of the order of \citep[cf.,][]{Ohira2013}
\begin{equation}
t_{\rm acc} \simeq \frac{p^2}{D_{TSA}} \simeq 10 \,\left(\frac{L_0}{\tau(p) c}\right)^{2/3}
                              \left(\frac{\langle \delta u^2 \rangle }{c^2}\right)^{-1} \tau(p)
                    \; \propto\; p^{\alpha/3}\,.
\end{equation} 
Comparing shear particle acceleration in (purely) turbulent flows to the one in non-relativistic laminar flows, 
i.e. $D_{sh}$ (eq.~[\ref{Dsh}]) and $D_{TSA}$ (eq.~[\ref{DTSA2}]), roughly yields $\frac{D_{TSA}}{D_{sh}} 
\simeq 30\, \frac{\langle \partial u^2 \rangle}{(\Delta u)^2} \left( \frac{\Delta r}{L_0}\right)^2 
\left(\frac{L_0}{\lambda_{\rm res}}\right)^{4/3}$ and suggests that for non-relativistic (!) flow velocities, shear 
acceleration in large-scale turbulent flows (assumed to be Kolmogorov-type) could be more effective than 
in laminar shear flows if the relevant shear transition region ($\Delta r$) is not sufficiently narrow. Note, 
however, that in reality the situation is more complex as many astrophysical flows exhibit some directionality 
(i.e., are composed of an underlying bulk velocity plus some turbulent fluctuations), so that an interplay 
between both effects may occur. An extension to relativistic turbulence still remains to be explored.

\section{Concluding remarks}
As described in this review, a variety of (not mutually exclusive) processes may be operative in astrophysical 
shear flows and facilitate particle transport and energization. Key results include the self-consistent generation 
of electromagnetic micro-turbulence and supra-thermal particle distributions as well as efficient Fermi-type 
particle acceleration in relativistic shearing flows. In particular, given sufficient turbulence, the latter processes 
can lead to a continued acceleration of charged particles, capable of producing power-law particle momentum
distributions as long as the velocity shear persists. This offers an interesting explanation for the extended 
high-energy emission observed in large-scale AGN jets. Similar processes can contribute to the energization 
of extreme cosmic rays. In general, injection of energetic seed particles (in particular with respect to electrons) 
and relativistic flow speeds are required for these processes to operate efficiently. In the case of AGN and 
GRBs the former condition can be met by first-order shock and/or classical second-order Fermi processes. 
For non-relativistic speeds, on the other hand, turbulent shear acceleration can be more efficient than shear 
acceleration in quasi laminar flows.\\ 
Open issues concerning our understanding of shear particle acceleration include extensions of PIC simulations 
to 3D and magnetized shear flows, a detailed characterization of the diffusive transport in fast shearing flows 
along with the reaction effects of accelerated particles, as well as a generalization of turbulent shear to relativistic 
velocity turbulences. It is hard to see, however, how velocity shear could not play a role in the energization of 
charged particles. High-resolution studies of astrophysical jets can offer complementary information concerning 
their internal structure and provide relevant constraints for more detailed applications.

\vspace{6pt} 
\funding{Funding by a DFG Heisenberg Fellowship RI 1187/6-1 is gratefully acknowledged.}

\acknowledgments{It is a pleasure to thank Peter Duffy and Felix Aharonian for collaboration over 
the years, and Martin Lemoine and Gary Webb for recent discussions on the topic. I also would like 
to thank Paulo Alves, Edison Liang, Kohta Murase and Yutaka Ohira for permission to use figures 
from their papers. Useful comments by the anonymous referees are gratefully acknowledged.}

\externalbibliography{yes}
\bibstyle{mdpi}
\bibliography{references}

\end{document}